\documentclass[%
 reprint,
superscriptaddress,
longbibliography,
 amsmath,amssymb,
aps,
 prb,
floatfix,
]{revtex4-2}

\usepackage[T1]{fontenc}
\usepackage[utf8]{inputenc}
\usepackage[english]{babel}
\usepackage{stackengine}
\usepackage[]{graphicx}
\usepackage[caption=false]{subfig}
\usepackage{amsmath}
\usepackage{amssymb}
\usepackage{amsfonts}
\usepackage[]{units}
\usepackage{times}
\usepackage{physics}
\usepackage{bm}
\usepackage{bbold}
\usepackage{xcolor}

\usepackage{tabularx}

\usepackage{hyperref}
\hypersetup{colorlinks=true, citecolor=blue, urlcolor=teal, linkcolor=blue}

\newcommand{\p}{\parallel}

\usepackage{letltxmacro}
\LetLtxMacro{\ORIGselectlanguage}{\selectlanguage}
\makeatletter
\DeclareRobustCommand{\selectlanguage}[1]{%
  \@ifundefined{alias@\string#1}
    {\ORIGselectlanguage{#1}}
    {\begingroup\edef\x{\endgroup
       \noexpand\ORIGselectlanguage{\@nameuse{alias@#1}}}\x}%
}
\newcommand{\definelanguagealias}[2]{%
  \@namedef{alias@#1}{#2}%
}

\makeatother

\definelanguagealias{en}{english}
\definelanguagealias{EN}{english}
\definelanguagealias{eng}{english}
\definelanguagealias{de}{ngerman}

\begin{document}

\title{Spatial Exciton Localization at Interfaces of Metal Nanoparticles and Atomically Thin Semiconductors}

\author{Robert Salzwedel}\email{r.salzwedel@tu-berlin.de}
\affiliation{Institut für Theoretische Physik, Nichtlineare Optik und Quantenelektronik, Technische Universität Berlin, Berlin, Germany}
\author{Lara Greten}
\affiliation{Institut für Theoretische Physik, Nichtlineare Optik und Quantenelektronik, Technische Universität Berlin, Berlin, Germany}
\author{Stefan Schmidt}
\affiliation{Institut für Theoretische Physik, Nichtlineare Optik und Quantenelektronik, Technische Universität Berlin, Berlin, Germany}
\author{Stephen Hughes}
\affiliation{Department of Physics, Queen's University, Kingston, Ontario, K7L 3N6,  Canada}
\author{Andreas Knorr}
\affiliation{Institut für Theoretische Physik, Nichtlineare Optik und Quantenelektronik, Technische Universität Berlin, Berlin, Germany}
\author{Malte Selig}
\affiliation{Institut für Theoretische Physik, Nichtlineare Optik und Quantenelektronik, Technische Universität Berlin, Berlin, Germany}

\begin{abstract}

We present a
self-consistent Maxwell-Bloch theory to analytically study the interaction between a nanostructure consisting of a metal nanoparticle and a monolayer of transition metal dichalcogenide. For the combined system, we identify an effective eigenvalue equation that governs the center-of-mass motion of the dressed excitons in a plasmon-induced potential. Examination of the dynamical equation of the exciton-plasmon hybrid reveals the existence of bound states with negative eigenenergies, which we interpret as excitons localized in the plasmon-induced potential. The appearance of these bound states in the potential indicates strong coupling between excitons and plasmons.
We quantify this coupling regime by computing the scattered light in the near-field explicitly and identify signatures of strong exciton-plasmon coupling with an avoided crossing behavior and an effective Rabi splitting of tens of meV.
\end{abstract}
\maketitle
\section{Introduction}

%
Transition metal dichalcogenides (TMDCs) exhibit remarkable optical properties, such as a direct bandgap in the monolayer limit \cite{mak2010atomically,splendiani2010emerging}, valley-selective dichroism \cite{cao2012valley,xiao2012coupled}, and a spin-split band structure \cite{xiao2012coupled,kosmider2013large}. These properties, along with their high sensitivity to the surrounding environment \cite{rytova2018screened,keldysh1979coulomb,feierabend2017proposal,greben2020intrinsic,rosner2016two}, make them ideal for functionalization \cite{benson2011assembly} with external nanoparticles, such as molecules \cite{katzer2023impact,christiansen2023optical,greben2020intrinsic,thompson2023optical}, metal nanoparticles \cite{pincelli2023observation,mueller2018microscopic,kusch2021strong}, quantum dots \cite{deng2023effective}, or other 2D materials \cite{geim2013van,rivera2015observation,ovesen2019interlayer}, to locally tailor their optical properties.

Recently, many research works have focused on localizing excitons in TMDC layers through the deterministic creation of defects within the structure \cite{barthelmi2020atomistic}, strain-induced localization that attracts carriers \cite{tonndorf2015single,he2015single}, and Moiré potentials \cite{seyler2019signatures,mahdikhanysarvejahany2022localized}. 
These methods enable the localization of individual excitons in TMDCs, making them promising candidates for single-photon emitters in 2D hybrid materials \cite{gao2023atomically,barthelmi2020atomistic}.
Moreover, the strong coupling of electromagnetic modes to quantum emitters provides unprecedented control over the quantum states, which may have applications, in particular when it reaches the quantum optics limit where light-matter effects cannot be explained semi-classically and  multi-photon correlations are essential \cite{li2022room,fernandez2018plasmon,franke2020quantized}.  

There has been much recent interest in creating joint states of excitonic and plasmonic excitations, known as {\it plexcitonic} states \cite{manjavacas2011quantum,sang2021tuning,xiong2021room,zhang2021steering};
these hybrid states can be observed in strongly interacting systems that support both excitons and plasmons. 
Most works have focused on systems where an excitonic system is located inside a cavity to take advantage of the local field enhancement 
\cite{deng2002condensation,deng2010exciton,kleemann2017strong,gross2018near}.
For example, picocavities have been utilized to achieve strong exciton-plasmon interaction \cite{gross2018near, kleemann2017strong,baumberg2019extreme}. Similarly, other systems have also reached the strong coupling regime \cite{zhu2023electroluminescence}, evidenced by a clear spectral splitting.
However, the past few years have witnessed a significant surge in interest towards a variety of systems, in particular, systems consisting of individual nanoparticles interacting with excitonic systems, which were shown to reach the strong coupling regime without requiring a typical dipole-cavity interaction commonly used in cavity-QED \cite{manjavacas2011quantum,sang2021tuning,xiong2021room,zhang2021steering}:
In particular, experiments revealed impressive Rabi splittings on the order of $\unit[100]{meV}$ \cite{han2018rabi,geisler2019single,stuhrenberg2018strong} for systems consisting of nanorods \cite{wen2017room}, resonators \cite{kleemann2017strong,qin2020revealing}, nanodisks \cite{abid2017temperature,geisler2019single}, bipyramids \cite{stuhrenberg2018strong} and nanocubes \cite{han2018rabi}. 

On the theoretical side, these interacting systems are mostly treated using a classical coupled mode theory with the interaction strength as fitting parameter \cite{wu2010quantum,torma2015strong}. However, there have been two recent studies that investigate the strong coupling of a metal nanorod with a TMDC monolayer \cite{denning2022quantum,carlson2021strong} based on a quasinormal mode analysis and quantum reaction coordinate approach, respectively, which reproduce the experimentally observed spectral splittings \cite{zhu2023electroluminescence,geisler2019single,stuhrenberg2018strong}. However, the modifications of the excitonic properties are so far not well investigated. \\

\begin{figure}[htb]
    \centering
    \includegraphics[width=\linewidth]{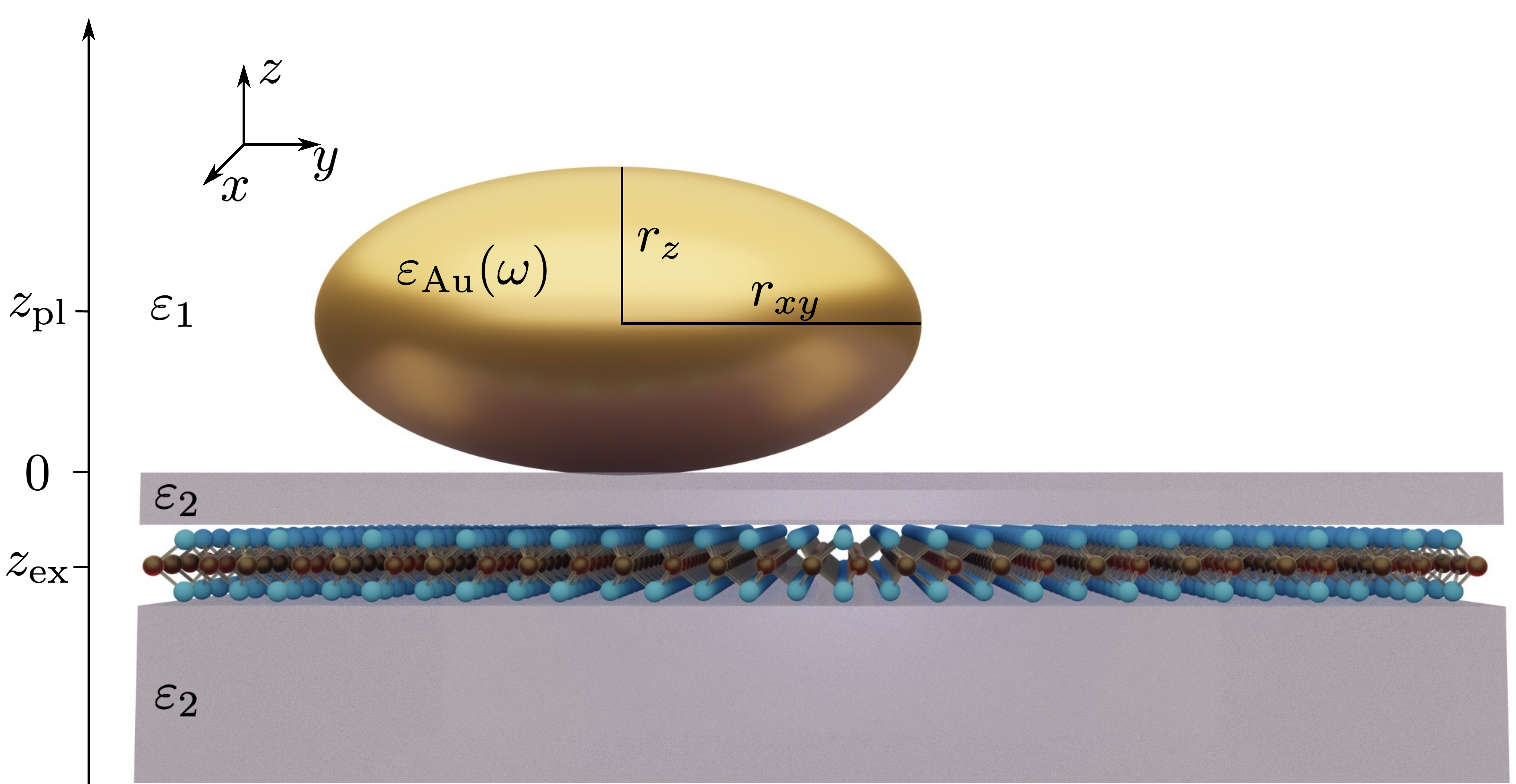}
    \begin{picture}(0,0)
    \put(22,65){\includegraphics[width = .43\linewidth]{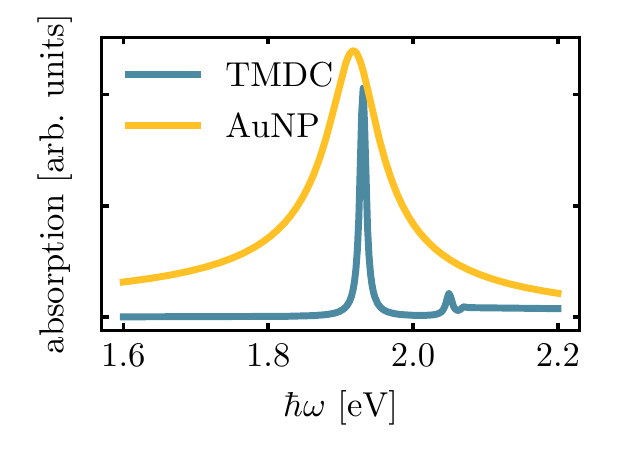}}
    \end{picture}
    \caption{\textbf{Coupled Nanostructure:} The investigated system is composed of a gold nanoparticle with its center being located at $z_{\text{pl}}$ and a two-dimensional TMDC monolayer at $z_{\text{ex}}$. The half spaces reveal a constant background permittivity $\varepsilon_1$/$\varepsilon_2$, respectively, so that an interface exist at $z=0$. The eccentricity of the AuNP can be manipulated from a sphere to an oblate spheroid which influences the interaction strength. The inset shows the absorption spectrum of the TMDC (blue) and the polarizability of a spheroidal gold nanoparticle. Using the permittivity function from Ref.~\cite{etchegoin2006analytic} and Mie theory \cite{mie1908beitrage,gans1912uber}, the polarizability can be calculated classically.
    }
    \label{fig:system}
\end{figure}

In this paper, we consider a hybrid nanostructure consisting of a spheroidal gold nanoparticle (AuNP) and a monolayer of transition metal dichalcogenides (TMDC), as illustrated in Fig.~\ref{fig:system}. 
To model this system in a quantitative way, both the TMDC and the AuNP are encapsulated in two different media with homogeneous and isotropic permittivities $\varepsilon_1$ and $\varepsilon_2$, respectively.
In contrast to previous works, we treat the entire system using a semi-classical microscopic model based on the Maxwell-Heisenberg equation of motion framework. We analytically identify an eigenvalue equation in the composite system, which describes the center-of-mass motion of the excitons in the potential induced by the plasmonic excitation. 
Our derived eigenvalue equation can be used to drastically reduce the numerical complexity of the problem, and offers new physical insight into the character of the hybridization. As an example, it allows us to connect to the strong coupling limit with the occurrence of bound exciton states induced by the AuNP.

The rest of our paper is organized as follows: In Sec.~\ref{sec:maxwell_bloch}, we give an introduction to the theoretical description of the composite subsystems: We provide the excitonic Bloch equation for TMDC excitons in Sec.~\ref{sec:excitons} and discuss Mie-Gans theory for the localized plasmons in the AuNP in Sec.~\ref{sec:plasmon}. We proceed with introducing a Green's function solution of Maxwell's equations in Sec.~\ref{sec:coupled_nanostructure} that is used to couple the constituents and find equations that describe the dynamics of the coupled system. 
In Sec.~\ref{sec:plexcitonicstates}, we discuss the occurring eigenvalue equation that characterizes the interaction within the nanoparticle and analyze the corresponding eigenvalues and eigenvectors.
In Sec.~\ref{sec:localization}, we then study the implications of the interaction and the arising eigenstates on the macroscopic polarization within the TMDC monolayer. Section~\ref{sec:strongcoupling} computes the electric near-field around the nanostructures and finds a peak splitting of the excitonic and plasmonic modes that for artificially detuned exciton resonance is shown to result in an avoided crossing behavior of the two resonances. 
Finally, in Sec.~\ref{sec:conclusion}, we provide our  conclusions and discuss the utility of our analytical plexcitonic approach to describe the interaction in nanostructures and their implications on localization and strong coupling. 


\section{Maxwell-Bloch approach}
\label{sec:maxwell_bloch}

We employ a semi-classical framework to obtain a set of self-consistent Maxwell-Bloch equations. These describe the exciton dynamics in terms of excitonic Bloch equations, the plasmonic response of the metal nanoparticle using Mie theory and the electromagnetic field that mediates the interaction by Maxwell's equations. 
We work in real space for the out-of-plane component ($z$ direction) and Fourier transform the in-plane component ($x$, $y$-directions) and transform the temporal dynamics to the frequency domain so that we use a set of $(\vb{Q}_\p, z ;\,\omega)$ as our coordinates, which makes it possible to solve the differential equations algebraically. 

\subsection{Optical response of TMDC excitons}
\label{sec:excitons}

The microscopic dynamics in TMDC monolayers are described using the Heisenberg equation of motion, which leads to excitonic Bloch equations, as described in Refs.~\cite{selig2019ultrafast,katsch2018theory,knorr1996theory}: 
\begin{align}
    \label{eq:blochequation}
	\biggl(E^{\nu} + \frac{\hbar^2\vb{Q_\parallel}^2}{2M} &-\hbar \omega - i\gamma^\nu  \biggr)
    \,p^{\xi\nu}_{\vb{Q_\p}}(\omega) 
    \\ \nonumber
	&= 
	\varphi^*_{\nu}(\vb{r_\p=0})\,
	(\vb{d}^{\xi})^*\cdot\vb{E}_{\vb{Q_\parallel}}( z_{\text{ex}};\,\omega).
\end{align}
The left-hand side accounts for the oscillation of the microscopic polarization, $p^{\xi\nu}_{\vb{Q_\p}}$, with excitonic energy, $E^\nu$, where we use a valley index $\xi = \pm 1$ for the $K/K'$ valley, respectively, exciton state number $\nu$, e.g., 1s, 2s, \dots, and Fourier component of the center-of-mass motion $\vb{Q_\p}$. 
Furthermore, the left-hand side accounts for the dispersion of excitons (second term) with the exciton mass $M$. 
The dephasing rates $\gamma^\nu$ are added to account for phonon-induced dephasing as calculated microscopically in Ref.~\cite{selig2016excitonic}. 
The TMDC excitons are driven by the electric field $\vb{E}_{\vb{Q_\p}} (z_{\text{ex}};\,\omega)$ via the electronic transition dipole moment $\vb{d}^{\xi }$ at the respective valley \cite{xiao2012coupled} and $\varphi_{\nu}(\vb{r_\p=0})$. 
Here, $\varphi_{\nu}(\vb{r_\p})$ is the excitonic wave function depending on the relative coordinate of electron and hole $\vb{r_\p}$ and can be obtained by solution of the Wannier equation with a Coulomb potential tailored to the specific geometry. 
In the literature, this is mostly done using a Rytova-Keldysh-type approach \cite{rytova2018screened,keldysh1979coulomb} with a model dielectric function \cite{cappellini1993model,trolle2017model}, which we adapt to our effective four layer system, cf.~Fig.~\ref{fig:system}. 
The excitonic wave function is evaluated at $\vb{r_\p=0}$ which accounts for the probability of finding electron and hole at the same position \cite{kira2006many}.
The resulting TMDC absorption spectrum is given in the inset in Fig.~\ref{fig:system}. 
In the hybrid structure, the total field at the TMDC position $\vb{E}_{\vb{Q_\p}}(z_{\text{ex}})$ includes the external field $\vb{E}_{\vb{Q_\p}}^0$, a contribution caused by the AuNP-TMDC interaction as well as the inter- and intra-valley exchange coupling within the monolayer \cite{qiu2015nonanalyticity}.

The microscopic TMDC polarization $p^{\xi\nu}_{\vb{Q_\p}}$, calculated from Eq.~\eqref{eq:blochequation}, is related to the macroscopic polarization via
\begin{align}
		\vb{P}^{\text{TMDC}}_{\vb{Q_\parallel}}(z;\,\omega)
		= \sum_{\xi\nu}
		\vb{d}^{\xi}\,
		&\varphi_{\nu}(\vb{r_\p=0})\,
		p^{\xi\nu}_{\vb{Q_\p}}(\omega)\,
		\delta(z-z_{\text{ex}})\nonumber\\ &+ \text{c.c.}, 
  \label{eq:TMDpolarization_basic}
\end{align}
where we assume that the monolayer can be approximated as infinitesimally thin which makes it effectively two dimensional. 
This definition will be used as a source term in the macroscopic Maxwell's equation. 
Later in this paper, we will only focus on the $\nu = \text{1s}$ resonance and effectively drop the $\nu$ index. This is a good approximation when the 1s resonance is spectrally clearly separated and the spectral range is limited to the one dominated by the 1s resonance as it is for our case.

\subsection{Optical response of nanoparticle plasmons}
\label{sec:plasmon}
%
The metal nanoparticle, in a dipole approximation, is modeled using Mie-Gans theory \cite{mie1908beitrage,gans1912uber} that condenses the light-matter interaction in response to the external field $\vb{E}_{\vb{Q_\p}}^0$ for a spheroid in a diagonal polarizability tensor, $\boldsymbol{\alpha}(\omega)$, whose diagonal components are given by: 
\begin{align}
    \label{eq:polarizability}
    \alpha_i(\omega) = 4\pi \varepsilon_0 \varepsilon_1 \frac{r_x r_y r_z}{3}\frac{\varepsilon_{\text{Au}}(\omega)-\varepsilon_1}{L_i \varepsilon_{\text{Au}}(\omega)+\varepsilon_1(1-L_i)}.
\end{align}
The gold permittivity $\varepsilon_{\text{Au}}(\omega)$ is analytically modeled using the approach from Ref.~[\onlinecite{etchegoin2006analytic}], that incorporates two interband transitions in the visible regime in order to accurately describe the experimental data found in Ref.~[\onlinecite{johnson1972optical}]. 
Its analytical expression is given in Eq.~\eqref{eq:permittivity}. 
%
%
The aspect ratio and such the lengths of the spheroid's semi-axes ($r_i$), determine the strength of the individual components via $L_i$, defined in Eqs.~\eqref{eq:lxly} and \eqref{eq:lz}.
The choice of an oblate spheroid allows for enhanced interaction of AuNP and TMDC since it reduces the effective separation while keeping the volume and thus the polarizability large. 
In the inset of Fig.~\ref{fig:system}, the absolute value of the in-plane polarizability of the considered spheroid is shown as an example. 
All used parameters can be found in Tab.~\ref{tab:table_parameters}. 

In this dipole approximation, the AuNP polarization can be written as
\begin{align}
    \vb{P}^{\text{AuNP}}_{\vb{Q_\p}}(z;\,\omega)
    =
    \frac{\boldsymbol{\alpha}(\omega)}{(2\pi)^2}\cdot 
    \int \dd^2\vb{Q_\p'}\,
        &e^{-i(\vb{Q_\p}-\vb{Q_\p'})\cdot \vb{r_\p^\text{pl}}}\,
        \vb{E}_{\vb{Q}'_\parallel}(z_{\text{pl}})\nonumber\\
        &\times\delta(z-z_\text{pl}),\label{eq:goldpolarization}
\end{align}
which describes the polarization of a point dipole located at $\vb{r}_{\text{pl}} = (\vb{r}_\p^{\text{pl}}, z_{\text{pl}})$. 
The polarizability $\boldsymbol{\alpha}(\omega)$ incorporates the electric field generated by the AuNP. Thus, $\vb{E}_{\vb{Q'_\parallel}}(z_{\text{pl}})$  corresponds to the electric field at the position of the AuNP, excluding the field contributed by itself.
For the purpose of this paper, we will assume $\vb{r}^{\text{pl}}_\p=0$.
Combing the two polarizations given in Eqs.~\eqref{eq:TMDpolarization_basic} and \eqref{eq:goldpolarization}, the full polarization is given by  
\begin{align}
    \vb{P}_{\vb{Q_\p}}(z;\,\omega) = \vb{P}^{\text{TMDC}}_{\vb{Q_\p}}(z;\,\omega) +\vb{P}^{\text{AuNP}}_{\vb{Q_\p}}(z;\,\omega),
\end{align}
which enters Maxwell's equations 
to compute the electric field close to the nanostructure.

\smallskip
\subsection{Optical response of the coupled nanostructures}
\label{sec:coupled_nanostructure}
In our description, the interaction of TMDC and AuNP is mediated by the electric field, as can be seen in Eqs.~\eqref{eq:blochequation} and \eqref{eq:goldpolarization}, which has to be determined self-consistently from Maxwell's equations. 
The starting point for the investigation is the wave equation
\begin{align}
\label{eq:waveequation}
    \qty(\bm{\nabla}^2 -\frac{\varepsilon(z)}{c^2}\pdv[2]{t})\vb{E}(\vb{r},t) = 
    &\frac{1}{\varepsilon_0 c^2}\pdv[2]{t}\vb{P}(\vb{r},t)\\\nonumber
    -&\frac{1}{\varepsilon(z)\varepsilon_0}\bm{\nabla} \qty (\bm{\nabla} \cdot\vb{P}(\vb{r},t)),
\end{align}
for polarization, $\vb{P}(\vb{r},t)$, in a background medium with spatially piecewise constant permittivity $\varepsilon(z)$, which is $\varepsilon_1$ in the upper half plane and $\varepsilon_2$ in the lower half plane.

A general solution of this equation can be provided via the Green's function using the coordinate system $(\vb{Q}_\p, z ;\,\omega)$ which we obtain by Fourier transformation ${\bf E}({\bf r};\,\omega) = \frac{1}{(2\pi)^2}\int\dd^2\vb{Q_\p} e^{i \vb{Q_\p}\cdot \vb{r_\p}} \vb{E}_{\vb{Q_\p}}(z;\,\omega)$, with
\begin{align}
    \vb{E}_{\vb{Q_\p}} (z;\,\omega) &= \int_\mathbb{R} \dd z' \mathcal{G}_{\vb{Q_\p}} (z,z';\,\omega)\cdot \vb{P}_{\vb{Q_\p}} (z';\,\omega) \nonumber \\
    &+ \vb{E}_{\vb{Q_\p}}^{0} (z;\,\omega),\label{eq:electricfield}
\end{align}
with the dyadic Green's function $\mathcal{G}_{\vb{Q_\p}} (z,z';\,\omega)$ and the external electric field $\vb{E}_{\vb{Q_\p}}^0 (z;\,\omega)$.
For Eq.~\eqref{eq:waveequation}, the dyadic Green's function is given by
\begin{align}
  & \mathcal{G}_{\vb{Q_\p}} (z,z';\,\omega) = \nonumber \\
    & \ \ \qty[ -\frac{\omega^2}{\varepsilon_0 c^2}\mathbb{1} 
    +\frac{1}{\varepsilon_0\varepsilon(z)}\mqty(
        \vb{Q}_\parallel \otimes \vb{Q}_\parallel
        & 
        i \vb{Q_\p}\partial_{z'}
        \\
        i \vb{Q}_\p^T \partial_{z'}
        & 
        \partial_{z'}^2
        )
        ]G_{\vb{Q}_\parallel}(z,z';\,\omega),
        \label{eq:full_dyadic_green}
\end{align}
where the symbol $\mathbb{1}$ denotes the three-dimensional identity matrix. 
The second matrix has a 2 by 2 matrix as its first entry, and the resulting matrix is also three-dimensional. 
Here, the scalar Green's function $G_{\vb{Q}_\parallel}(z,z';\,\omega)$ is defined as
\begin{align}
    G_{\vb{Q}_\parallel}(z,z';\omega) = -\frac{i}{2k_{\vb{Q_\p}}}e^{ik_{\vb{Q_\p}} \abs{z-z'}}, 
    \label{eq:full_scalar_green}
\end{align}
where $k_{\vb{Q_\p}} \equiv \sqrt{\varepsilon(z)\frac{\omega^2}{c^2}-Q_\p^2}$. 
Eqs.~\eqref{eq:full_dyadic_green} and \eqref{eq:full_scalar_green} allow one to calculate the self-consistent electric field at the TMDC and the AuNP position which enters the dynamical equation for the microscopic TMDC polarization $p^{\xi\nu}_{\vb{Q_\p}}(\omega)$, cf.~Eq.~\eqref{eq:blochequation}, and the AuNP polarization, cf.~Eq.~\eqref{eq:goldpolarization}. 

We will focus on the 1s TMDC resonance in our interacting system, and for clarity, we will omit the index $\nu$. We will use the notation $\varphi^{\text{1s}}(\vb{r_\p}=0)\rightarrow\varphi_0$ to represent the value of $\varphi^{\text{1s}}$ at the origin and the corresponding 1s damping coefficient $\gamma^{\text{1s}}$.
By inserting Eq.~\eqref{eq:electricfield} into Eq.~\eqref{eq:blochequation}, we find the following equation of motion for the microscopic TMDC polarization:
\begin{widetext}
\begin{align}
    \qty[
        E^{\text{1s}} + \frac{\hbar^2 \vb{Q}_\parallel^2}{2M} - \hbar \omega- i \gamma
    ]
    p_{\vb{Q}_\parallel}^\xi(\omega)
    =
    \varphi_0^*\,
    \vb{d}^{\xi *}
    \cdot 
    \biggl[
    \vb{E}_{\vb{Q}_\parallel}^0 &(z_{\text{ex}};\,\omega) 
    + \mathcal{G}_{\vb{Q_\p}} (z_{\text{ex}},z_{\text{pl}};\,\omega)\cdot
    \boldsymbol{\alpha}(\omega)\cdot
        \vb{E}^0(\vb{r}_{\text{pl}};\,\omega)
    \biggr]\label{eq:GreensEOM}\\
     +\abs{\varphi_0}^2
    \vb{d}^{\xi *}
    \cdot
    \sum_{\xi'}
    \biggl[
        \mathcal{G}_{\vb{Q_\p}} (z_{\text{ex}},z_{\text{ex}};\,\omega)\cdot
        \vb{d}^{\xi'} p_{\vb{Q}_\parallel}^{\xi'}(\omega)
        &+
        \mathcal{G}_{\vb{Q_\p}} (z_{\text{ex}},z_{\text{pl}};\,\omega)\cdot
        \frac{\boldsymbol{\alpha}(\omega)}{(2\pi)^2}\cdot\int\dd^2\vb{Q_\p'}\,
            \mathcal{G}_{\vb{Q_\p'}} (z_{\text{pl}},z_{\text{ex}};\,\omega)\cdot
            \vb{d}^{\xi'}
            p_{\vb{Q}'_\parallel}^{\xi'}(\omega)
    \biggr]\nonumber.
\end{align}

\end{widetext}
In Eq.~\eqref{eq:GreensEOM}, the coupling between TMDC excitons and the AuNP plasmon induced by the electric field is given in terms of the Green's functions, including the self-interaction of the excitonic polarization. 
In Eq.~\eqref{eq:GreensEOM}, the first term on the right-hand side is the interaction with the external electric field $\vb{E}_{\vb{Q}_\parallel}^0(z_{\text{ex}})$ at the TMDC position $z_{\text{ex}}$.
The second term is the external electric field at the AuNP position, which is resonantly enhanced by the AuNP and then coupled to the TMDC. 
In the second line, we see that the electric field also mediates a dipole-dipole coupling between the excitons at the $K/K'$ point, widely known as the inter- and intra-valley exchange coupling \cite{glazov2014exciton}. %
The final term in the equation describes a self-interaction of the TMDC that is mediated by the AuNP, as evidenced by the appearance of two Green's functions. This term can be interpreted as an effective exciton-exciton interaction.


In our particular setup, special care is required to include the dielectric interface at $z=0$, which arises due to the piecewise constant background permittivity.
Since the distance between the TMDC and the AuNP is only a few nanometers and the wavelengths used are in the optical range, we have opted to utilize the quasi-static Green's function, provided in Eq.~\eqref{eq:greenfunctionstatic}, which also incorporates the change in background permittivity.
This leads to the fact that the quasi-static Green's function can only be defined piecewise. 
Due to the interface, the Green's function also contains additional mirror charge terms. 
The Green's function is derived following Ref.~\cite{de2010optical} and takes into account the individual positions of the scatterers. 

In the quasi-static limit, i.e., $c \rightarrow \infty$, the dyadic Green's function can be expressed as
\begin{align}
    \label{eq:greenfunctionstaticdyadic}
    \mathcal{G}^{\text{st}}_{\vb{Q_\p}} (z,z') 
    = 
     \ \ \frac{1}{\varepsilon_0 \varepsilon(z)}\mqty(
        \vb{Q}_\parallel\otimes\vb{Q}_\parallel
        & 
        i \vb{Q_\p}\partial_{z'}
        \\
        i \vb{Q}_\p^T \partial_{z'}
        & 
        \partial_{z'}^2 
        ) G^{\text{st}}_{\vb{Q}_\parallel}(z,z').
\end{align}
Evaluating Eq.~\eqref{eq:GreensEOM} with the quasi-static scalar Green's function $G^{\text{st}}_{\vb{Q}_\parallel}(z,z')$ in Eq.~\eqref{eq:greenfunctionstatic}, we obtain individual equations for the respective valley $K/K'$.
To investigate the effects resulting from the coupling of TMDC and AuNP, we first diagonalize our system of equations by performing a transformation with respect to the inter-valley exchange coupling (first term in the second line in Eq.~\eqref{eq:GreensEOM}):
\begin{align}
    \label{eq:diagonalization}
	\mqty(
		p_{\vb{Q_\p}}^U\\
		p_{\vb{Q_\p}}^V
		)
		\equiv
	\frac{1}{\sqrt{2}}
	\mqty(-e^{i\phi}&e^{-i\phi}\\
			e^{i\phi}&e^{-i\phi}
		)
	\cdot 
	\mqty(
		p_{\vb{Q_\p}}^K\\
		p_{\vb{Q_\p}}^{K'}
		),
\end{align}
similar to Ref.~\cite{qiu2015nonanalyticity} with $\phi$ being the angle coordinate in polar coordinate corresponding to $\vb{Q_\p}$. 
The same matrix transformation is used to transform the circularly polarized external electric field $\vb{E}_{\vb{Q_\p}}^0$ in Eq.~\eqref{eq:GreensEOM} into its new basis $\biggl\{E_{\vb{Q_\p}}^{0,U}$, $E_{\vb{Q_\p}}^{0,V}\biggr\}$. 
We find two decoupled equations, Eqs.~\eqref{eq:undisturbed} and \eqref{eq:diagonalizedmbe}, for the new polarizations $p_{\vb{Q_\p}}^U(\omega)$ and $p_{\vb{Q_\p}}^V(\omega)$:
\begin{align}
\label{eq:undisturbed}
    &\biggl[
        E^{\text{1s}} + \frac{\hbar^2 \vb{Q_\p^2}}{2M}-\hbar \omega -i\gamma
    \biggr]
    p_{\vb{Q_\p}}^{U}(\omega) = 
    d^* \varphi^*_0\ E_{\vb{Q_\p}}^{0,U}(z_{\text{ex}};\,\omega).
\end{align}

In Eq.~\eqref{eq:undisturbed}, $p_{\vb{Q_\p}}^U$ is unaffected by the exchange coupling. The left-hand side of Eq.~\eqref{eq:undisturbed} exhibits a free parabolic exciton dispersion that is consistent with previous literature \cite{qiu2015nonanalyticity}. 
Accordingly, we will refer to Eq.~\eqref{eq:undisturbed} as parabolic Bloch equation.
It's worth noting that both the exchange coupling contributions and the coupling contributions between TMDC and AuNP cancel each other out. This is due to the quasi-static approach, which reduces the interaction to longitudinal components that appear under the transformation in Eq.~\eqref{eq:diagonalization} only in the $V$ component, cf.~Ref.~\cite{qiu2015nonanalyticity}.
Hence, the right-hand side only accounts for the excitation by the external electric field $E_{\vb{Q_\p}}^{0,U}(z_{\text{ex}};\,\omega)$ at the TMDC position and accordingly has the same form as the pristine TMDC case without exchange and TMDC-AuNP coupling. 
In contrast, the equation for $p_{\vb{Q_\p}}^V(\omega)$, Eq.~\eqref{eq:diagonalizedmbe}, reads:
\begin{align}
\label{eq:diagonalizedmbe}
	\biggl[
		E^{\text{1s}} + \frac{\hbar^2 \vb{Q_\p^2}}{2M}&+X_{\vb{Q}_\p}(z_{\text{ex}})-\hbar \omega - i\gamma
	\biggr]
	p_{\vb{Q_\p}}^{V}(\omega)\\\nonumber
	-&\frac{1}{(2\pi)^2}\int\dd^2\vb{Q_\p'} \ V_{\vb{Q}_\p\vb{Q'_\p}}(z_{\text{ex}},z_{\text{pl}};\,\omega)
	\ p_{\vb{Q'_\p}}^{V}(\omega)\\\nonumber
	=& d^*\varphi^*_{0}\ \biggl(E_{\vb{Q_\p}}^{0,V}(z_{\text{ex}};\,\omega)
	\ + S_{\vb{Q_\p}}(z_{\text{pl}},z_{\text{ex}};\,\omega)\biggr).
\end{align}
Comparing to Eq.~\eqref{eq:undisturbed}, where all interaction contributions cancel, we find three additional terms. 
The first one is the intra- and inter-valley exchange term, which renormalizes the parabolic dispersion: 
\begin{align}
    X_{\vb{Q}_\parallel}(z_{\text{ex}}) = -\abs{\varphi_0}^2 \abs{d}^2 \frac{Q_\parallel^2}{\varepsilon_0\varepsilon_2} G_{\vb{Q}_\parallel}^{\text{st}} (z_{\text{ex}},z_{\text{ex}}).
\end{align}

\begin{figure}
    \centering
    \includegraphics[width = \linewidth]{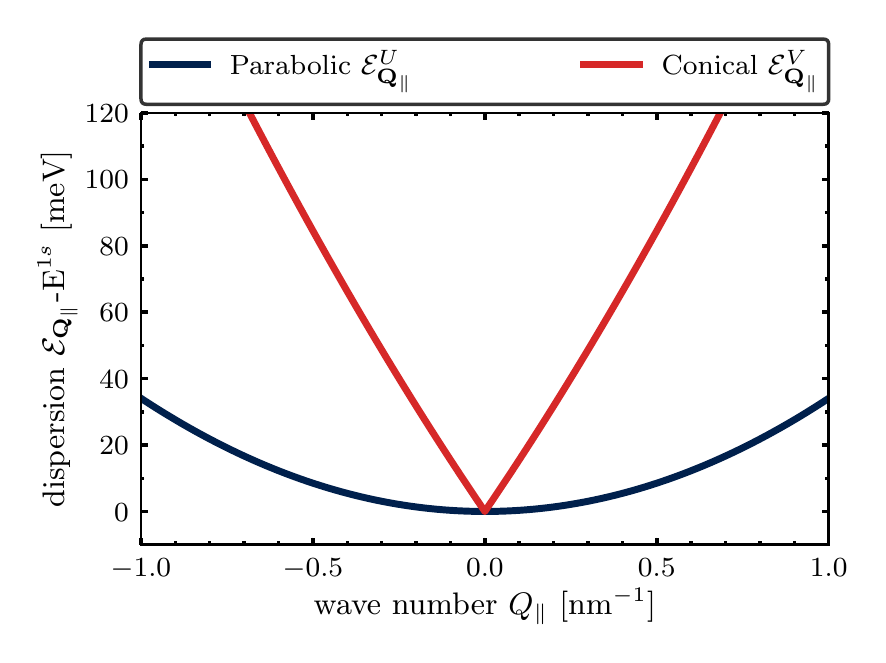}
    \caption{\textbf{Dispersion:} The exchange coupling causes a split in the dispersion, resulting in parabolic dispersion for the non-interacting $p_{\vb{Q_\p}}^U$, and a conical dispersion for $p_{\vb{Q_\p}}^V$, which experiences interaction with the gold nanoparticle. The distinctive shape of the dispersion has also influenced its nomenclature.}
    \label{fig:dispersion}
\end{figure}

As can be seen in Fig.~\ref{fig:dispersion}, where we depict the further relevant momentum range from \unit[-1]{nm$^{-1}$} to \unit[1]{nm$^{-1}$}, $X_{\vb{Q_\p}}(z_{\text{ex}})$ changes the parabolic dispersion to a conical one depending on the exchange coupling among the $K/K'$ valleys, cf.~Ref.~\cite{qiu2015nonanalyticity}.
Hence, we refer to Eq.~\eqref{eq:diagonalizedmbe} as the conical Bloch equation.
The other additional terms, $V_{\vb{Q}_\p\vb{Q'_\p}}(z_{\text{ex}},z_{\text{pl}};\,\omega)$ and $S_{\vb{Q_\p}}(z_{\text{pl}},z_{\text{ex}};\,\omega)$, are given by:
\begin{widetext}
\begin{align}
    V_{\vb{Q}_\parallel\vb{Q}'_\parallel} (z_{\text{ex}},z_{\text{pl}};\,\omega)
    &= 
    \abs{\varphi_0}^2 \abs{d}^2
    \frac{Q_\parallel^2}{\varepsilon_0\varepsilon_2}     G_{\vb{Q}_\parallel}^{\text{st}} (z_{\text{ex}},z_{\text{pl}})
    \frac{{Q'_\parallel}^2}{\varepsilon_0\varepsilon_1} G_{\vb{Q}'_\parallel}^{\text{st}}(z_\text{pl},z_{\text{ex}})
    \qty[
        \alpha_\p(\omega)\cos(\phi-\phi')
        + \alpha_z(\omega) 
        ],
    \label{eq:potential}\\
    S_{\vb{Q}_\parallel}(z_{\text{ex}},z_{\text{pl}};\,\omega) &=\frac{Q_\p^2}{\varepsilon_0\varepsilon_2}
    G_{\vb{Q}_\parallel}^{\text{st}} (z_{\text{ex}},z_{\text{pl}})
        \qty[
            \alpha_\p(\omega) E_0^{V}(\vb{r}_{\text{pl}};\omega)
            - i \alpha_z(\omega) E_0^z(\vb{r}_{\text{pl}};\omega)
        ]\label{eq:source}.
\end{align}
\end{widetext}

The first of the two terms, $V_{\vb{Q}_\parallel\vb{Q}'_\parallel}$, describes effects of the effective exciton-exciton interaction mediated by the plasmonic nanoparticle. 
This has the form of coupling between induced dipoles, as apparent from the characteristic cosine dependence on the relative angle $\phi-\phi'$.
In the following, it is interpreted as an additional potential for the center-of-mass motion of the excitons. Due to the symmetry of the system, we chose $\alpha_\parallel = \alpha_x = \alpha_y$.
The term on the right-hand side, $S_{\vb{Q_\p}}(z_{\text{pl}})$, represents the excitation caused by the external electric field. This excitation is initially scattered and enhanced by the AuNP before coupling to the TMDC.
The interaction mediated via the in-plane and the $z$-axis of the AuNP, respectively, is qualitatively different, as can be seen from the additional imaginary unit in front of the $z$-component.

In agreement with Ref.~\cite{qiu2015nonanalyticity}, we show in Fig.~\ref{fig:dispersion} that the exchange coupling in Eq.~\eqref{eq:diagonalizedmbe} leads to the formation of a parabolic lower band and a conical upper band in the excitonic dispersion. 
For this reason, we have chosen $U$ and $V$ as indices for the parabolic and conical dispersion, respectively.
We define their dispersion from
\begin{align}
    \mathcal{E}_{\vb{Q}_\p}^U &= E^{\text{1s}} + \frac{\hbar^2 \vb{Q_\p^2}}{2M},\label{eq:dispersion_excitonic}\\
    \mathcal{E}_{\vb{Q}_\p}^V &= E^{\text{1s}} + \frac{\hbar^2 \vb{Q_\p^2}}{2M}+X_{\vb{Q}_\p}(z_{\text{ex}}).\label{eq:dispersion_plexcitonic}
\end{align}

Summarizing our analytical advances so far, we have found that our equations can be diagonalized such that only one of the TMDC exciton components, Eq.~\eqref{eq:diagonalizedmbe}, is affected by the AuNP, while the other component, Eq.~\eqref{eq:undisturbed}, is completely unchanged compared to the purely excitonic case in Eq.~\eqref{eq:blochequation}. 
%

\section{Plexcitonic States}
\label{sec:plexcitonicstates}

Similar to identifying the Wannier equation in the semiconductor Bloch equation \cite{lindberg1988effective,haug2004quantum,kira2006many} which captures the relative motion of electron and hole. In the conical Bloch equation, Eq.~\eqref{eq:diagonalizedmbe}, which captures the modification of the dispersion due to exchange coupling as well as the interaction of AuNP plasmon and TMDC excitons, we identify the following eigenvalue equation that describes the full excitonic center-of-mass motion with in-plane momentum $\vb{Q_\p}$:
\begin{align}
    \label{eq:eigenvalue}
	\Biggl[
		\frac
			{\hbar^2 \vb{Q_\p^2}}
			{2M}
		+X_{\vb{Q}_\p}(z_{\text{ex}})
	\Biggr]
	&\Psi_{\vb{Q_\p}}^{\text{R},\lambda}\\\nonumber
	-\frac{1}{(2\pi)^2}\int \dd^2\vb{Q'_\p}\
	&V_{\vb{Q}_\p\vb{Q'_\p}}(z_{\text{ex}},z_{\text{pl}};\,\omega)\
	\Psi_{\vb{Q_\p'}}^{\text{R},\lambda}
	= 
	E^\lambda\Psi_{\vb{Q_\p}}^{\text{R},\lambda},
\end{align}
%
%
The non-local plasmon-induced potential $V_{\vb{Q}_\p\vb{Q}'_\p}$ determines the center-of-mass motion $\vb{Q_\p}$ on the dispersion modified by the exchange coupling (left side in Eq.~\eqref{eq:eigenvalue}). 
Although the Wannier equation and the plexcitonic eigenvalue equation,  Eq.~\eqref{eq:eigenvalue}, which we treat as a Schrödinger equation, share formal similarities, they differ qualitatively because the plasmon-induced potential $V_{\vb{Q_\p}\vb{Q_\p'}}$ is complex due to the complex-valued polarizability $\boldsymbol{\alpha}(\omega)$, cf.~Eq.~\eqref{eq:potential}. 

Accordingly, the eigenvalue equation, Eq.~\eqref{eq:eigenvalue}, becomes non-Hermitian which results in complex-valued eigenvalues and requires to distinguish left and right eigenvectors $\Psi_{\vb{Q_\p'}}^{\text{L},\lambda}$ and $\Psi_{\vb{Q_\p'}}^{\text{R},\lambda}$ \cite{moiseyev2011non,gilary2005calculations} as will be done in Sec.~\ref{sec:eigenvectors}.
We will refer to these new eigenstates as \textit{plexcitonic states} as they describe the hybridized plasmon-exciton states of plasmonic and excitonic character. 

In this section, we study the eigenvalue equation numerically and analyze the eigenvalues and eigenvectors in detail which we will use in subsequent sections to define macroscopic quantities. For this purpose, we choose an oblate spheroid as depicted in Fig.~\ref{fig:system}. The explicit parameters can be found in Tab.~\ref{tab:table_parameters}. 

\subsection{Eigenvalues of hybrid structure}
\label{sec:eigenvalues}

The eigenvalue analysis of Eq.~\eqref{eq:eigenvalue} by numerical eigendecomposition in analogy to established methods for the Wannier equation \cite{berghauser2014analytical,katsch2018theory} reveals a finite number of eigenvalues with negative real part representing bound states (discussion below). 
The eigenvalues with positive real part distribute quasi-continuously along $\mathcal{E}_{\vb{Q}_\p}^{U/V}$. 
Figure~\ref{fig:dispersion} shows the dispersion: $\mathcal{E}_{\vb{Q}_\p}^{V}$ is conical for the parameter range of interest, consistent with recent work \cite{qiu2015nonanalyticity}. 
For increasing background permittivity $\varepsilon_2$, the dispersion interpolates between a cone and a parabola.
Further discussion on this behavior can be found in App.~\ref{sec:eigenvalues_appendix}.

\begin{figure}[htb]
    \centering
    \includegraphics{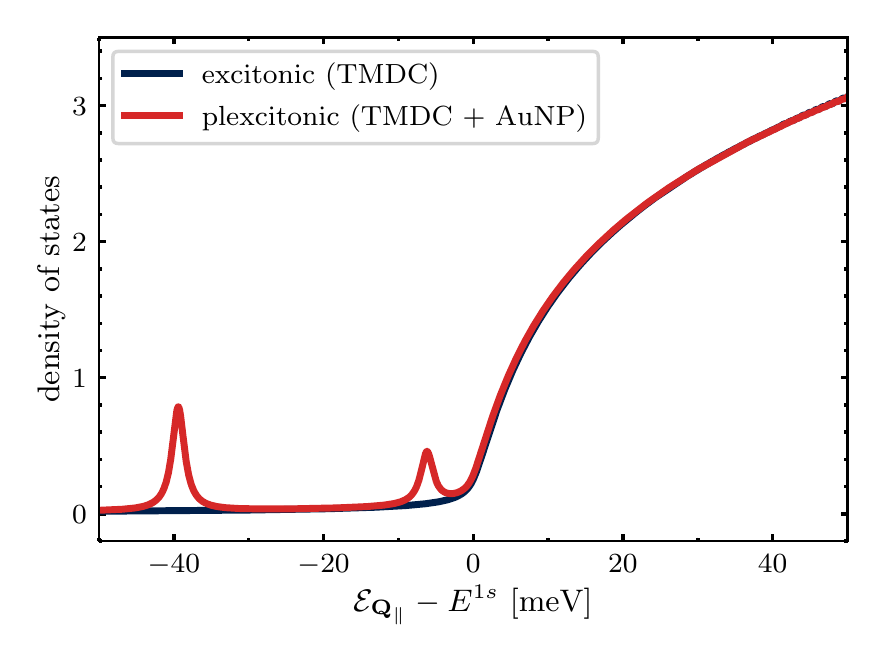}
    \caption{\textbf{Excitonic Density of States.} Comparison of the excitonic density of states for the interacting plexcitonic and the purely excitonic system where the potential $V_{\vb{Q_\p}\vb{Q_\p'}}$ in Eq.~\eqref{eq:eigenvalue} was set to zero artificially for a spheroid with $r_{xy} = \unit[8]{nm}$ and $r_z = \unit[4]{nm}$. The plexcitonic plot exhibits two additional peaks at negative energy, corresponding to the coupling via the in-plane component $\alpha_\p$ of the AuNP, with a multiplicity of two, and the z component $\alpha_z$ with a multiplicity of one. The peak at $\approx \unit[-39]{meV}$ corresponds to the in-plane coupling, while the one at $\approx \unit[-8]{meV}$ is caused the interaction via the $z$ component. For the graphical representation, we used $\gamma_\ell = $\unit[1]{meV}.}
    \label{fig:DOS}
\end{figure}

Through a parameter study of the background permittivities $\varepsilon_1$ and $\varepsilon_2$, the aspect ratio of the ellipsoid $\flatfrac{r_{xy}}{r_{z}}$, and the distance between AuNP and TMDC $\abs{z_{\text{pl}}-z_{\text{ex}}}$, we observe up to three eigenvalues with negative real part up to \unit[100]{meV} as well as associated eigenvectors (discussion below). The imaginary contribution (broadening in the spectrum) is on the same order of magnitude. 
These eigenvalues correspond to an attractive interaction mediated by the plasmon-induced potential $V_{\vb{Q_\p}\vb{Q}_\p'}$ in Eq.~\eqref{eq:eigenvalue} that spatially localize excitons. 
We found that each of these eigenvalues originates from the interaction with the plasmonic mode along one of the three Cartesian axes of the nanoparticle. 
The frequency dependence of the binding energies is discussed in App.~\ref{sec:eigenvalues_appendix}. 
To illustrate these results, we calculate the excitonic density of states (DOS),
\begin{align}
    \text{DOS}(E) = \frac{1}{A}\sum_{\lambda}\delta(E-E^\lambda),
    \label{eq:DOS}
\end{align}
by evaluating the Dirac delta distribution, $\delta(E-E^\lambda)$, for the real part of the eigenvalues only.
In order to be able to plot the DOS, we approximate the delta distribution with Lorentzian functions $\mathcal{L}_{\gamma_\ell}(E,E^\lambda)$, which introduces an artificial linewidth $\gamma_\ell$.
Using $\gamma_\ell = \unit[1]{meV}$, 
Fig.~\ref{fig:DOS} shows that only the eigenvalues with negative real parts deviate from the quasi-continuous spectrum.
%
By switching the interaction with the external particle on and off in our numerical implementation, we can compare the purely excitonic system to the interacting plexcitonic one that includes the effective exciton-exciton interaction $V_{\vb{Q_\p Q_\p'}}$, mediated via the plasmonic nanoparticle. 
For our choice of parameters (oblate spheroid), cf.~Tab.~\ref{tab:table_parameters}, we find two interaction-induced peaks at negative energies that results in a non-vanishing density of states at the respective eigenvalue energy. 
Due to the symmetry of the spheroid, we find that the eigenvalues corresponding to the interaction via the in-plane axes are degenerate and cause the peak at $\unit[-39]{meV}$, while the peak at $\unit[-8]{meV}$ originates from interaction via the out-of-plane AuNP axis. A detailed parameter study for which parameters we obtain negative eigenvalues and localized eigenstates is provided in App.~\ref{sec:eigensystem_appendix}. 

Figure~\ref{fig:DOS} displays the excitonic DOS for the conical excitonic dispersion without interaction with the nanoparticle, represented by the blue line. The real and positive eigenvalues are quasi-continuously distributed among the dispersion $\mathcal{E}_{\vb{Q}_\p}^V$. In contrast to strictly two-dimensional systems with parabolic dispersion, the DOS is not a step function due to the presence of a linear term in the dispersion relation, cf.~Eq.~\eqref{eq:dispersion_plexcitonic}, that depends on the center-of-mass momentum $\vb{Q_\p}$.

For the plexcitonic case (red), a numerical analysis shows that all eigenvalues with positive real part distribute on the conical dispersion and have negligible imaginary parts (on the order of the numerical accuracy). However, the eigenvalues with negative real part deviate significantly from the conical excitonic case, as seen in Fig.~\ref{fig:DOS}, and have non-negligible imaginary parts.
The imaginary parts of the eigenvalues originate from nature of the lossy plasmon resonance and Förster-type processes between TMDC exciton and AuNP plasmon and introduce additional dephasing channels \cite{katzer2023impact,thompson2023optical}.

Based on our findings, we can conclude that the plasmon-mediated exciton-exciton interaction leads to the formation of plexcitonic states, exhibiting negative real part of the eigenvalue. 
We interpret this feature as the formation of bound states, where the real part of the eigenvalue represents the binding energy. These states cause the deviation in the density of states from the conical excitonic case in Fig.~\ref{fig:DOS}.

\subsection{Eigenvectors and probability density}
\label{sec:eigenvectors}

In this subsection, we analyze the eigenvectors corresponding to the negative eigenvalues presented in the previous section. 
In the usual excitonic picture, solutions of the Wannier equation \cite{selig2016excitonic} describe the relative electron-hole motion and their wave functions represent the probability amplitudes of their motion.
%
%
Due to its non-Hermitian nature, the physical interpretation of the plexcitonic eigenvalue equation is not straightforward. It generates left and right eigenvectors 
$\Psi_{\vb{Q}_\p}^{L,\lambda}$, $\Psi_{\vb{Q}_\p}^{R,\lambda}$.
To address this issue, we follow the approach presented in Ref.~[\onlinecite{barkay2001complex}] 
and define the probability density, 
\begin{align}
\label{eq:probabilitydensity}
    \rho^\lambda(\vb{r_\p}) \equiv \Psi^{L,\lambda}(\vb{r_\p})\Psi^{R,\lambda}(\vb{r_\p}),
\end{align}
where we use a normalization scheme:
\begin{align}
\braket{\Psi_{\vb{Q}_\p}^{L,\lambda}}{\Psi_{\vb{Q}_\p}^{R,\mu}} = \delta^{\lambda\mu},
\label{eq:biorthonormality}
\end{align} 
for the left and right eigenvectors $\Psi_{\vb{Q}_\p}^{L,\lambda}$, $\Psi_{\vb{Q}_\p}^{R,\lambda}$ with the scalar product defined as a 2D momentum integral over $\vb{Q_\p}$. 

\begin{figure}[htb]
    \centering
    \includegraphics{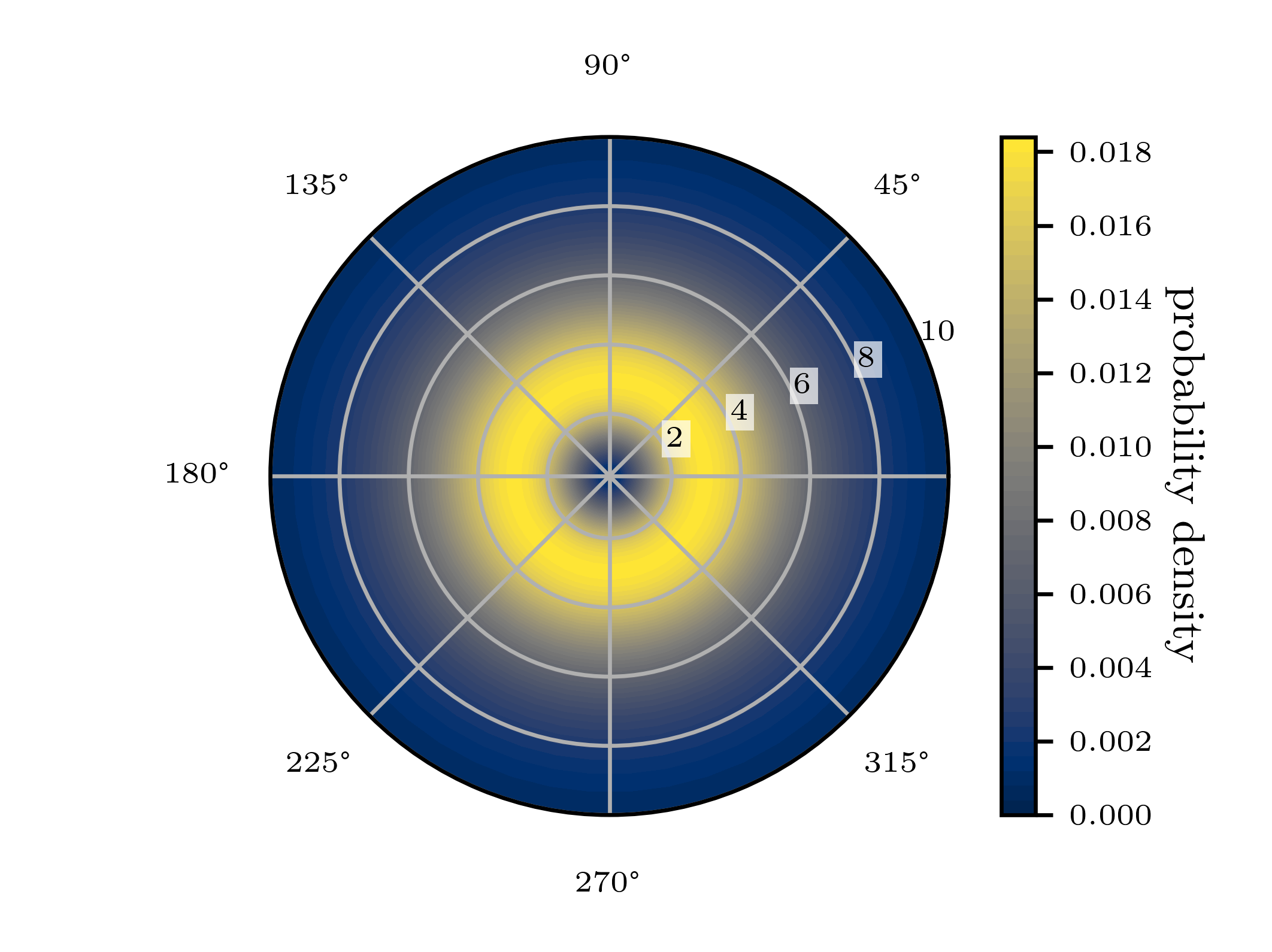}
    \begin{picture}(0,0)
    \put(-130,0){\includegraphics[width = .5\linewidth]{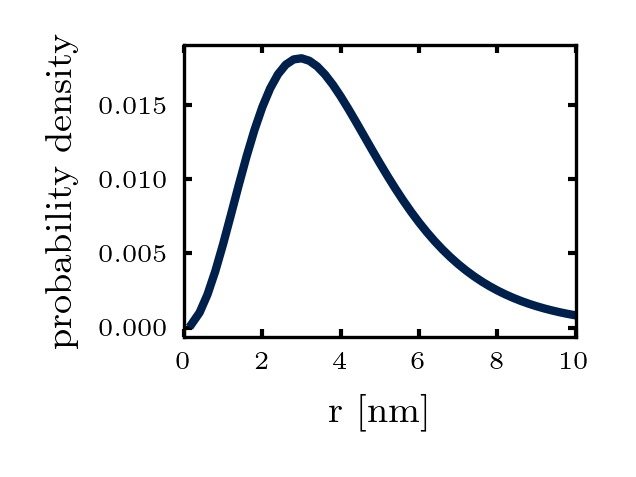}}
    \end{picture}
    \caption{\textbf{Probability density in real space.} The eigenvalues corresponding to in-plane interaction are degenerate, here we plot the superposition of the two probability densities corresponding to this attractive in-plane interaction. The inset shows the radial profile of the probability density.}
    \label{fig:eigenstates}
\end{figure}

Our analysis reveals that the eigenvectors $\Psi_{\vb{Q}_\p}^{L,\lambda}$ and $\Psi_{\vb{Q}_\p}^{R,\lambda}$, respectively, belonging to the three negative eigenvalues (bound states), correspond to the degeneracy of the spatial axes of the gold nanoparticle polarizability $\boldsymbol{\alpha}(\omega)$. 
They accurately reflect the symmetry of the coupling axis, showing either an $x$- or $y$-orientation or a radial symmetry for coupling via the out-of-plane component. 
As expected, we observe that these eigenvectors related to negative eigenvalues are localized near the origin, thus representing bound states, while the eigenvectors corresponding to positive eigenvalues are spread out throughout momentum space and represent the discretization of the considered Hilbert space. 
Therefore, the AuNP allows to study exciton localization in the vicinity of the AuNP. 
To illustrate this, we discuss the real space probability density $\rho(\vb{r_\parallel})$, defined in Eq.~\eqref{eq:probabilitydensity}.
In Fig.~\ref{fig:eigenstates}, we plot the real part of the sum of the probability densities, cf.~Eq.~\eqref{eq:probabilitydensity}, associated with the degenerate eigenvalue from the in-plane coupling, resulting in a ring-shaped distribution around the origin. The $x$ and $y$ components individually exhibit orientation along their respective axes.

The ring-shaped feature is a result of the in-plane dipole-dipole interaction between the spatially fixed dipole (plasmon) and the dipole that is free to move in a 2D plane (exciton). The interaction that results from dipole-dipole interaction via the $z$ component of the plasmon is studied in App.~\ref{sec:eigenvectors_appendix}.

In summary, we find that the additional states in the density of states reveal negative eigenenergies, cf.~Fig.~\ref{fig:DOS}. These states are spatially confined near the gold nanoparticle, indicating that they correspond to bound states.
In Sec.~\ref{sec:localization} \& \ref{sec:strongcoupling}, we will analyze the implications of these states on macroscopic observables such as the macroscopic TMDC polarization and the electric near-field in more detail.

\smallskip
\smallskip
\smallskip
\smallskip
\section{Localization}
\label{sec:localization}

In this section, we analyze the impact of the external nanoparticle on the macroscopic polarization within the TMDC layer, contributing to optical observables via Maxwell's equations. As a macroscopic observable, we use the absolute value of the TMDC polarization in Eq.~\eqref{eq:TMDpolarization_basic}, that we map on the plexcitonic eigenstates defined by Eq.~\eqref{eq:eigenvalue}, using the plexcitonic expansion 
\begin{align}
    \label{eq:plexcitonicexpansion}
    p_{\vb{Q_\p}}^V(\omega)=\sum_\lambda \Psi_{\vb{Q_\p}}^{\text{R},\lambda}\,p^\lambda(\omega),
\end{align}
with suitable expansion coefficients $p^\lambda$. We expand Eq.~\eqref{eq:diagonalizedmbe} using the plexcitonic expansion in Eq.~\eqref{eq:plexcitonicexpansion}, expressing it in terms of right eigenvectors $\Psi_{\vb{Q_\p}}^{R,\lambda}$ which form a complete basis in momentum space. We then project this expanded equation onto the corresponding left eigenvectors $\Psi_{\vb{Q_\p}}^{L,\lambda}$ and utilize the biorthonormality relation [see Eq.~\eqref{eq:biorthonormality}].

This approach yields a dynamical equation for the expansion coefficient $p^\lambda$, which we term the {\it plexcitonic polarization equation}:
\begin{align}
\label{eq:excitationequation}
	p^\lambda(\omega)= \frac{d^* \varphi^*_{0} }{(2\pi)^2}
	\int\dd^2\vb{Q'_\p}
		\frac{\qty(\Psi_{\vb{Q'_\p}}^{L,\lambda})^* \qty(E_{\vb{Q'_\p}}^{0,V}(z_{\text{ex}})
            +S_{\vb{Q'_\p}})}{E^{\text{1s}}+E^\lambda-\hbar \omega- i\gamma}.
\end{align}
We observe that the plexcitonic polarization $p^\lambda$ can be excited by two external source terms: the external field at the position of the TMDC, $E_{\vb{Q'_\p}}^{0,V}(z_{\text{ex}}) $, and the field scattered by the AuNP, $S_{\vb{Q'_\p}}$, as described in Eq.~\eqref{eq:source}. The latter carries a non-vanishing in-plane momentum $\vb{Q_\parallel}$. To simplify the notation, we no longer explicitly mention the dependencies of $S_{\vb{Q'_\p}}$. Notably, the complex-valued plexcitonic eigenvalues $E^\lambda$ renormalize not only the resonance energy, as seen in the denominator, but also the dephasing of the nanostructure, through their imaginary part which are negative and thus increase the effective dephasing of the nanostructure.

By Fourier transformation of Eq.~\eqref{eq:TMDpolarization_basic}, we find that the macroscopic TMDC polarization, including all contributions from $p_{\vb{Q}_\p}^U(\omega)$ and $p_{\vb{Q}_\p}^V(\omega)$, can be written as
\begin{widetext}
    \begin{align}
    \label{eq:TMDCpolarization}
    \vb{P}^\pm_{\text{TMDC}}(\vb{r};\,\omega) =&\frac{\abs{d}^2\abs{\varphi_0}^2}{2}
        \frac{1}{(2\pi)^2}\int \dd^2\vb{Q_\p}
            \Biggl\{
            e^{i\vb{Q_\p}\cdot \vb{r_\p}}
                \biggl[
                    \frac{1}{E^{\text{1s}} +\frac{\hbar^2\vb{Q_\p}^2}{2M}-\hbar \omega- i\gamma}
                    \mqty(
                        1 &  -e^{-2i\phi}\\
                        -e^{2i\phi} & 1
                        )\cdot
                    \vb{E}_{\vb{Q_\p}}^{0,\pm}
                        (z_{\text{ex}})
                        \\\nonumber
                    &+\frac{1}{(2\pi)^2}\int \dd^2\vb{Q_\p'}\sum_{\lambda} 
                        \frac{\Psi_{\vb{Q_\p}}^{R,\lambda}\,\qty(\Psi^{L,\lambda}_{\vb{Q_\p'}})^*}{E^{\text{1s}}+E^\lambda-\hbar \omega- i\gamma}
    			         \mqty(
                            e^{-i\phi}e^{i\phi'} & e^{-i\phi}e^{-i\phi'}\\
                            e^{i\phi}e^{i\phi'} & e^{i\phi}e^{-i\phi'}
                            )\cdot
                        \vb{E}_{\vb{Q_\p'}}^{0,\pm}
                        (z_{\text{ex}})\\\nonumber
        &+\frac{1}{(2\pi)^2}\int \dd^2\vb{Q_\p'}\sum_{\lambda}
        \frac{Q_\p^2}{\varepsilon_0\varepsilon_2}
    	\frac{
    		\Psi_{\vb{Q_\p}}^{R,\lambda}\qty(\Psi_{\vb{Q_\p'}}^{L,\lambda})^*    G_{\vb{Q}_\parallel}^{\text{st}}
    		}{E^{\text{1s}}+
    		E^\lambda-\hbar\omega-i\gamma
    	}
    		\mqty(
    			e^{-i\phi}e^{i\phi'} & e^{-i\phi}e^{-i\phi'}	&	-e^{-i\phi}	\\
    			e^{i\phi}e^{i\phi'} & e^{i\phi}e^{-i\phi'}	&	-e^{i\phi}		
    		)
    		\cdot
    		\mqty(
    			\alpha_{\p}\,E_0^{+}(\vb{r}_{\text{pl}})\\
    			\alpha_{\p}\,E_0^{-}(\vb{r}_{\text{pl}})\\
    			i\sqrt{2}\,\alpha_z	E_0^z(\vb{r}_{\text{pl}})
    		)
      \biggr] + {\rm c.c.}
    \Biggr\}.
    \end{align}
\end{widetext}

In Eq.~\eqref{eq:TMDCpolarization}, we can identify three distinct contributions to the macroscopic TMDC polarization. The first term corresponds to half the unperturbed response of the TMDC, which is independent of any interaction with the AuNP. The second term captures the interaction between the TMDC and the AuNP, as well as the TMDC self-interaction which is described by the plexcitonic eigenvalues $E^\lambda$ and eigenvectors $\Psi_{\vb{Q_\p}}^{L,\lambda}$ and $\Psi_{\vb{Q_\p}}^{R,\lambda}$. The third term represents the external electric field scattered and enhanced by the AuNP and subsequently transferred to the TMDC position where it contributes to the TMDC polarization. In the limit of vanishing AuNP, the third term vanishes and the second one reproduces the second half of the unperturbed TMDC response. 

\begin{figure}[h!]
    \centering
    \includegraphics{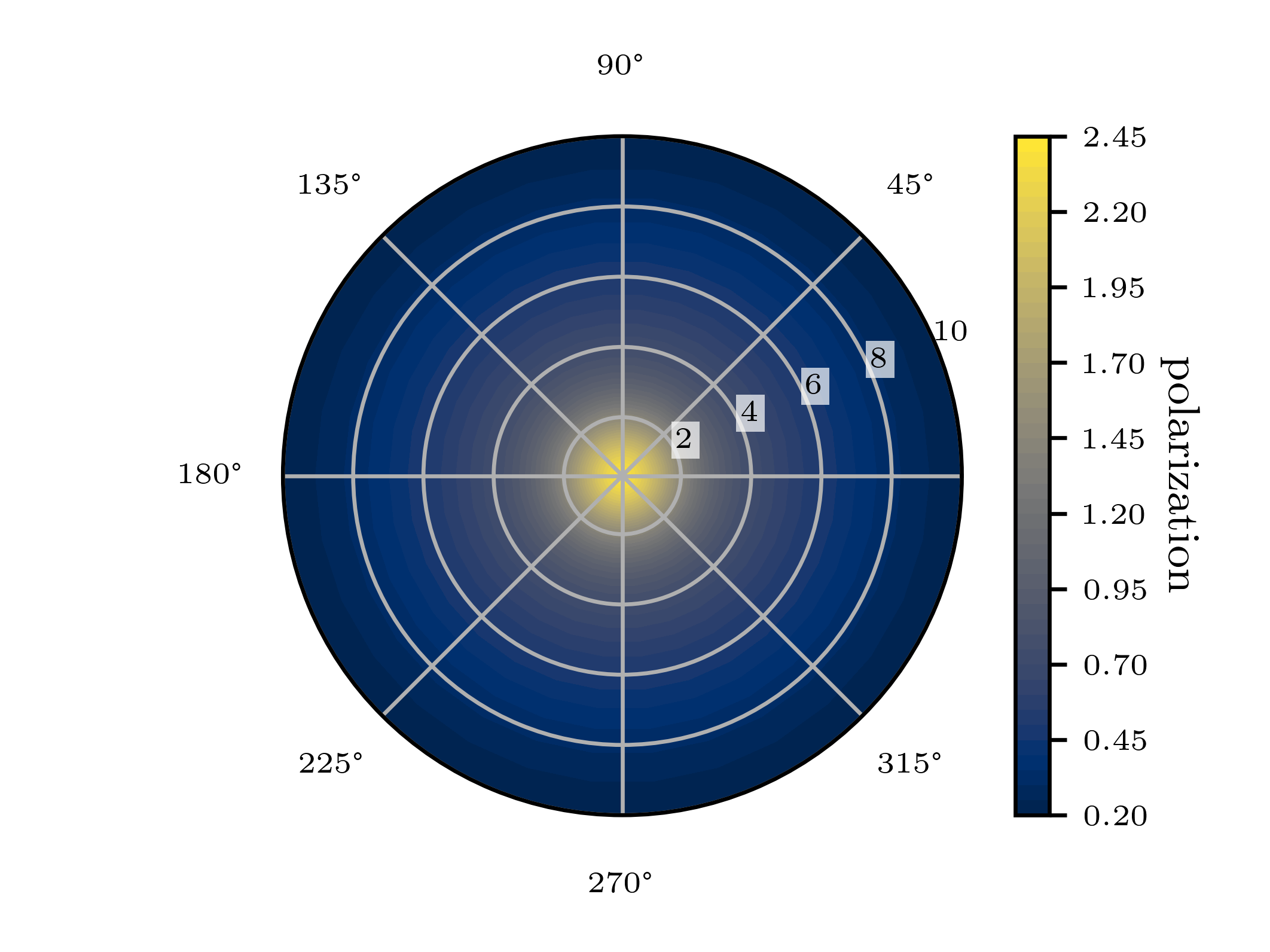}
    \begin{picture}(0,0)
    \put(-130,0){\includegraphics[width = .5\linewidth]{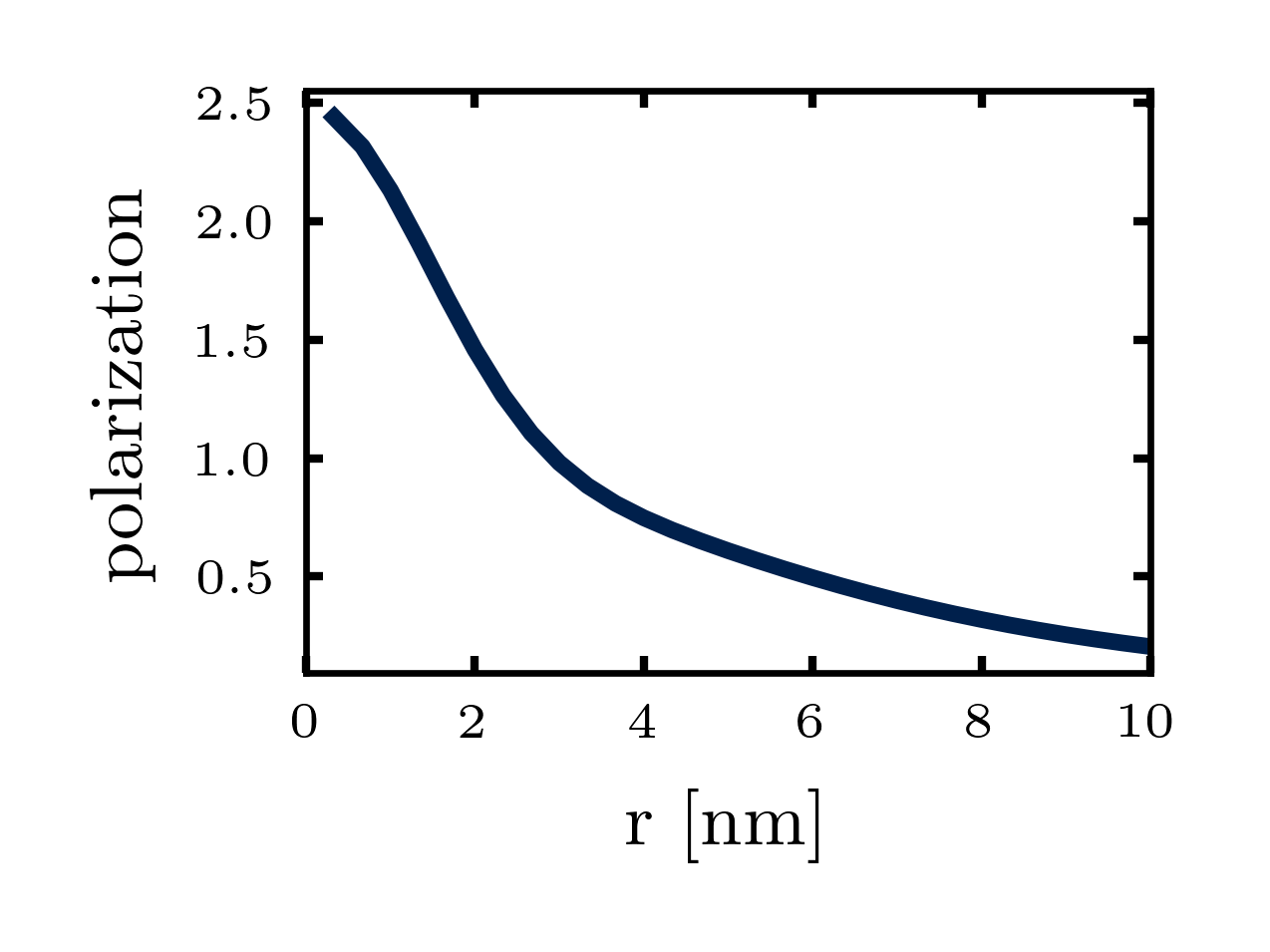}}
    \end{picture}
    \caption{\textbf{TMDC polarization in real space:} In the figure, we see for $\sigma^+$ excitation that full TMDC polarization localizes in a radially symmetric way below the metal nanoparticle.}
    \label{fig:polarization}
\end{figure}

%
In Fig.~\ref{fig:polarization}, we plot the resulting macroscopic polarization of the TMDC from Eq.~\eqref{eq:TMDCpolarization} when excited by a $\sigma^+$-polarized plane wave and find a radially symmetric distribution of the polarization around the nanoparticle location. 
According to Eq.~\eqref{eq:TMDCpolarization}, the spatial localization is mainly attributed to two key processes. The first one is the dipole-dipole interaction between the TMDC exciton and the AuNP plasmon, leading to the localized plexcitonic states discussed in Sec.~\ref{sec:plexcitonicstates}. The second one is the locally enhanced electric field in the TMDC layer, which occurs due to the scattering of the field by the AuNP. These processes are discussed individually in App.~\ref{sec:TMDC_polarization_appendix}.

This analysis reveals that proper selection of parameters, positioning a gold nanoparticle on a TMDC monolayer can induce the formation of plexcitonic states through dipole-dipole interactions. The resulting polarization enhancement in the TMDC underneath the nanoparticle effectively localizes carriers in the vicinity of the AuNP.


\section{Strong coupling}
\label{sec:strongcoupling}

This section focuses on the study of the electric field emitted by the nanostructure in response to an external electric field. To calculate the electric field outside the nanostructure, we use the Green's method to solve the wave equation described in Eq.~\eqref{eq:waveequation}. We find for the electric field distribution surrounding the nanostructure:
\begin{align}
\label{eq:field}
    \vb{E}_{\vb{Q_\p}} (z;\,\omega) &= 
        \mathcal{G}_{\vb{Q_\p}}^{\text{st}} (z,z_{\text{ex}})\cdot 
        \vb{P}_{\vb{Q_\p}}^{\text{TMDC}} (z_{\text{ex}};\,\omega)\\ \nonumber
        &+ \mathcal{G}_{\vb{Q_\p}}^{\text{st}}(z,z_{\text{pl}})\cdot 
        \vb{P}_{\vb{Q_\p}}^{\text{AuNP}} (z_{\text{pl}};\,\omega)\\\nonumber
        &+\vb{E}_{\vb{Q_\p}}^{0} (z;\,\omega).
\end{align}
The TMDC polarization is defined in Eq.~\eqref{eq:TMDCpolarization} and the gold polarization can be found in App.~\ref{sec:gold_polarization}. 
Importantly, our analysis in Sec.~\ref{sec:maxwell_bloch} relies on the quasi-static approximation, which accurately describes the electric near-field where $\vb{Q_\p}\neq 0$ is dominant.  Therefore, our analysis is limited to the electric near-field, which is well-captured by our approach. Note, that to accurately describe the electric far-field and account for radiative processes, it would be necessary to include the $\vb{Q_\p}=0$ case in the calculation.
Numerical evaluation of the electric near-field from Eq.~\eqref{eq:field} yields an optical near-field spectrum.

\begin{figure}
    \includegraphics[width=\linewidth]{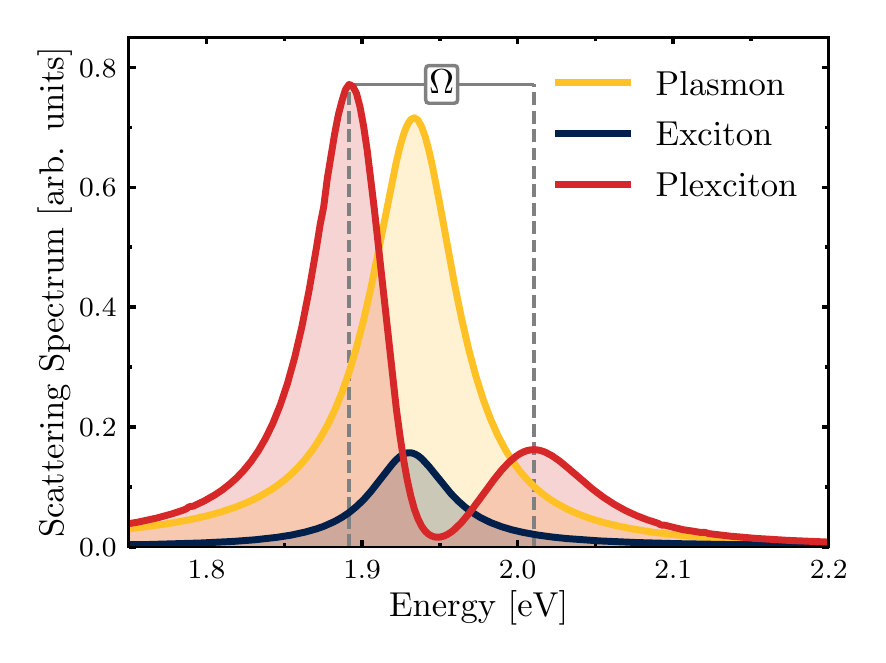}
    \caption{\textbf{Peak Splitting:} 
    For comparable excitonic and plasmonic resonance energies the response of the joint system splits into two distinct peaks at lower and higher energy compared to the shared resonance. The scattered electric field is detected at $x=\unit[2]{nm}$, $y=\unit[0]{nm}$, and $z=\unit[-5]{nm}$. The excitonic resonance energy is $E^{1s}= \unit[1.93] {eV}$, other parameters for room temperature can be found in Tab.~\ref{tab:table_parameters}. The plasmonic and plexcitonic spectra are presented in the correct ratio, the excitonic spectrum is scaled for display in the same plot.
    }

    \label{fig:peaksplitting}
\end{figure}

In Fig.~\ref{fig:peaksplitting}, we plot the Fourier transformed (purely real space) absolute value of the electric field intensity $\abs{\vb{E}(\vb{r}; \omega)}^2$ for excitation by plane waves. 
In contrast to TMDC excitation with a plane wave that only has a vanishing in-plane momentum, scattering off the AuNP generates electric field components in the near-field that possess a non-vanishing center-of-mass momentum $\vb{Q_\p}\neq 0$. These components can interact with momentum-dark excitonic states $\vb{Q_\p}\neq 0$ in the TMDC illustrating that the observed features result from dark excitons.
The spectra in Fig.~\ref{fig:peaksplitting} show that the individual non-interacting energy transitions of TMDC exciton and AuNP plasmon are designed so that their respective resonances, excitonic and plasmonic, occur at the same spectral location, cf.~Tab.~\ref{tab:table_parameters}, as depicted in the individual plasmon/exciton plots presented in Fig.~\ref{fig:system}.
However, for both systems in contact we observe spectral peak splitting, which is a sign of strong coupling between the individual TMDC exciton and AuNP plasmon oscillators \cite{carlson2021strong,denning2022quantum}.
Since our description relies on the excitation of dark excitons in the near-field, we attribute the occurrence of strong coupling to the spatial localization of near-field excited dark excitons.

\begin{figure}
    \centering
    \includegraphics[width=\linewidth]{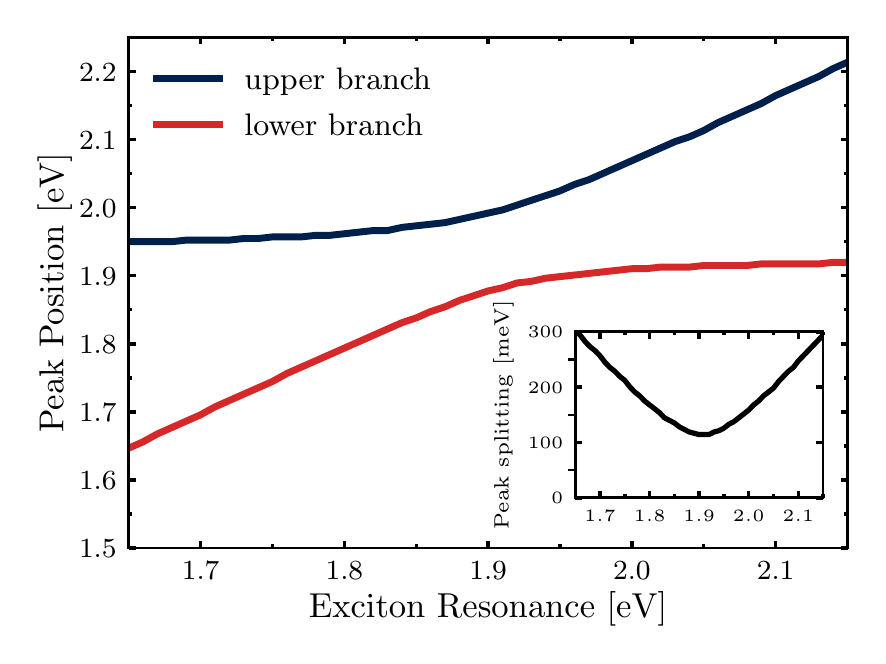}
    \caption{\textbf{Avoided crossing:} Numerically tuning the excitonic resonance allows to investigate the peak splitting in the plexcitonic spectrum and shows the avoided crossing behavior of the system indicating a strong-coupling regime. The given spectrum displays the peak positions, cf.~Fig.~\ref{fig:peaksplitting}, over a wide exciton resonance range. The inset illustrates the peak separation with its lowest value being the Rabi splitting that is approximately $\unit[110]{meV}$ in this case. All parameters used in the numerical implementation can be found in Tab.~\ref{tab:table_parameters}.}
    \label{fig:strongcoupling}
\end{figure}

We also highlight that our numerical approach allows us to artificially tune the excitonic resonance while keeping the plasmonic resonance fixed. Figure~\ref{fig:strongcoupling} shows the two peak positions with varying 1s excitonic resonance. 
Resonance energies far away from each other have little influence on one another, while we observe a significant peak splitting once the spectral separation of their peaks approaches their linewidths. Compared to the uncoupled case, the interaction leads to a minimum value of the spectral splitting of the observed spectral peaks, which we call effective Rabi splitting. This can be interpreted as avoided crossing behavior and supports the finding that the system behaves in a strong coupling regime.

\section{Conclusion and Perspectives}
\label{sec:conclusion}

We have presented a self-consistent theoretical approach for the near-field optical interaction between a monolayer of TMDC and a gold nanoparticle. 
Starting from the excitonic and plasmonic picture, we identified a novel eigenvalue equation that describes the center-of-mass motion of the excitons in an effective potential that features hybridized exciton-plasmon states. 
In this context, strong coupling is related to the excitation of momentum-dark excitons and their spatial localization in the monolayer near the AuNP: the density of states contains bound states below the excitonic 1s resonance. %
This interpretation is supported by the plexcitonic probability density and its influence on the spectral and spatial properties of the macroscopic TMDC polarization. 

Our analysis shows that the strong localization near the external particle leads to a strong coupling behavior visible in the electric near-field. 
Through a detailed parameter study, we establish a connection between the existence of these common states and an avoided crossing behavior in the spectral representation of the system. 
Our findings provide evidence that metal nanoparticles can be used to effectively localize excitons in two-dimensional TMDC layers.

\section{Acknowledgments}
%
We acknowledge fruitful discussions with Chelsea Carlson (Queen's University) as well as Manuel Katzer, Dominik Christiansen, and Jonas Grumm (TU Berlin).
This work is supported financially by the Deutsche Forschungsgemeinschaft (DFG) through Project SE 3098/1-1 (Project No. 432266622) and SFB 951 (Project No. 182087777).
We also acknowledge support from
the Natural Sciences and Engineering Research Council of Canada, and
the 
Alexander von Humboldt Foundation through a Humboldt Research Award.

\section*{Author Declarations}

The authors have no conflicts to disclose.
\appendix

\begin{table}[htb]
\centering
 \caption{Material parameters for TMDC and AuNP used in the numerical implementation }
 \begin{tabularx}{\linewidth}{Xllc}
 \hline\hline
    Parameter & Value &Unit& Reference \\
    \hline
    $d$      & $0.27$       & eC nm                                       & \cite{xiao2012coupled}\\
    $M$           &  $6.2535$        & fs$^2$ eV nm$^{-2}$     &   \cite{kormanyos2015theory}\\
    $\hbar\gamma$ (\unit[300]{K})  &     $0.0269$         & eV                        & \hyperlink{selig}{$^{\text{b}}$} \\
    $c$             &    $299.79246$    &  nm fs$^{-1}$                       &  \\
   $\varphi_0$       &     $0.51$  & nm$^{-1}$ & \hyperlink{rytova}{$^{\text{a}}$}\\
   \hline
   $\epsilon_\infty$  &  $1.53$ & &\cite{etchegoin2006analytic}\\
   $\hbar \omega_p$ & $8.55063$ &eV &\cite{etchegoin2006analytic}\\
   $\hbar \gamma_p$ & $0.072932$ & eV &\cite{etchegoin2006analytic}\\
   $A_1$ & $0.94$ &&\cite{etchegoin2006analytic}\\
   $\varphi_1$ & $-\pi/4$ &&\cite{etchegoin2006analytic}\\
   $\hbar \omega_1$ &  $2.64923$ & eV &\cite{etchegoin2006analytic}\\
   $\hbar \gamma_1$ & $0.53906$ & eV&\cite{etchegoin2006analytic}\\
   $A_2$ & $1.36$ &&\cite{etchegoin2006analytic}\\
   $\varphi_2$ & $-\pi/4$ &&\cite{etchegoin2006analytic}\\
   $\hbar \omega_2$ &$3.74575$&eV&\cite{etchegoin2006analytic}\\
   $\hbar \gamma_2$ &$1.31898$& eV&\cite{etchegoin2006analytic}\\
   \hline
   $r_{xy}$ &$8$& nm&\\
    $r_{z}$ &$4$& nm&\\
    $z_{0}$ &$-1$& nm&\\
    $z_{\text{pl}}$ &$5$& nm&\\
    $\varepsilon_1$ &$4.5$& &\\
    
    $\varepsilon_2$ &$1$& &\\
   \hline\hline
 \end{tabularx}\label{tab:table_parameters}
\flushleft
{
    \footnotesize
    {
        \hypertarget{rytova}{{$^{\text{a}}$}} Calculated from Rytova approach for 4 layer system similar to Ref.~\cite{rytova2018screened} \\
        \hypertarget{selig}{{$^{\text{b}}$}} Calculated by exploiting the method from Ref.~\cite{selig2016excitonic} 
    }
}
\end{table}

\section{Analytical model of the optical response of a gold nanoparticle}

We model the gold permittivity $\varepsilon(\omega)$ using the analytical expression provided in Ref.~\cite{etchegoin2006analytic}. It reads 
\begin{align}
    \label{eq:permittivity}
    \varepsilon_{\text{Au}}(\omega) = 
    \varepsilon_\infty
    &-\frac{\omega_p^2}{\omega(\omega + i \gamma_p)}\\\nonumber
    &+\sum_{j = 1,2}
        A_j\omega_j
        \biggl[
            \frac{e^{i\varphi_j}}{\omega_j -\omega -i\gamma_j}
            +
            \frac{e^{-i\varphi_j}}{\omega_i +\omega +i\gamma_j}
        \biggr],
\end{align}
where the first two terms describe a standard Drude model and the last terms additional interband transitions at the respective energies. 
The parameter values are given in Tab.~\ref{tab:table_parameters} and were obtained by the authors of Ref.~\cite{etchegoin2006analytic} as fits to the experimental data in Ref.~\cite{johnson1972optical}.

In order to incorporate the geometry of the nanoparticle, we employ Mie-Gans theory \cite{mie1908beitrage,gans1912uber} as we allow the AuNP to be spheroidal and obtain the AuNP polarizability given in Eq.~\eqref{eq:polarizability}. 
The impact of the aspect ratio is contained in the depolarization factors, $L_i$, which changes the optical response via the respective semi-axes, 
\begin{align}
L_x =L_y &= \frac{1}{2e_0^2}\qty(\frac{\sqrt{1-e_0^2}}{e_0}\arcsin(e_0)-(1-e_0^2)),\label{eq:lxly}\\
\text{and}\quad L_z &= \frac{1}{e_0^2}\qty(1-\frac{\sqrt{1-e_0^2}}{e_0}\arcsin(e_0)).\label{eq:lz}
\end{align}

For the oblate spheroid the $x$ and $y$ component coincide due to symmetry.
The eccentricity $e_0$ is defined to be
\begin{align}
e_0&=1-\frac{r_z^2}{r_{xy}^2},
\end{align}
and features the lengths of the semi-axes $r_i$.

\section{Screened potential}
\label{sec:rytova_potential}

For our model structure, we use a Rytova-Keldysh type approach \cite{rytova2018screened,keldysh1979coulomb} in order to calculate the potential that is later used in the Wannier equation in order to incorporate substrate effects on the carrier localization. 
The potential for our effective 4 layer system of $\varepsilon_1$ - $\varepsilon_2$ - TMDC - $\varepsilon_2$ reads:
\begin{align}
    V_{\vb{k}} = 
    \frac{q}{2\varepsilon_0\Tilde{\varepsilon} k}
        \frac{e^{2kL}+e^{kL}
        \qty(
            \delta_{\Tilde{\varepsilon},\varepsilon_2}
            +\delta_{\Tilde{\varepsilon},\varepsilon_2,\varepsilon_1})
            +\delta_{\Tilde{\varepsilon},\varepsilon_2}
            \delta_{\Tilde{\varepsilon},\varepsilon_2,\varepsilon_1}}
        {e^{2kL}-\delta_{\Tilde{\varepsilon},\varepsilon_2}
            \delta_{\Tilde{\varepsilon},\varepsilon_2,\varepsilon_1}},
\end{align}
with the definitions
\begin{align}
     \delta_{\Tilde{\varepsilon},\varepsilon_2} &= \frac{\Tilde{\varepsilon}-\varepsilon_2}{\Tilde{\varepsilon}+\varepsilon_2},\\
    \delta_{\Tilde{\varepsilon},\varepsilon_2,\varepsilon_1} &= \frac{\Tilde{\varepsilon}-\varepsilon_2 \xi}{\Tilde{\varepsilon}+\varepsilon_2\xi}, \qquad \xi = \frac{e^{2kR}-\delta_{\varepsilon_2,\varepsilon_1}}{e^{2kR}+\delta_{\varepsilon_2,\varepsilon_1}},
\end{align}
where $L$ is the thickness of the TMDC layer and R is the thickness of the intermediate layer between the TMDC and the $z=0$ plane, cf.~Fig.~\ref{fig:system}.\\

\section{Green's function}
\label{sec:greens_function}

For our specific geometry depicted in Fig.~\ref{fig:system}, including the interface of two background permittivities $\varepsilon_1$ and $\varepsilon_2$ and assuming the TMDC to be effectively two dimensional, the Green's function can be calculated. We find, in agreement with Refs.~\cite{de2010optical,jackson1999classical},
\begin{widetext}
\begin{align}
	G_{\vb{Q_\parallel}}^{\text{st}}(z,z') = 
	\begin{cases}
		-\frac{1}{2 Q_\parallel}
		e^{-Q_\parallel\abs{z-z'}}
		-\frac{1}{2Q_\parallel}
		\frac{\varepsilon_1-\varepsilon_2}{\varepsilon_1+\varepsilon_2}
		e^{-Q_\parallel\abs{z+z'}}
		&,\quad z,z' > 0 
		\\
		-\frac{1}{Q_\parallel}
		\frac{\varepsilon(z)}{\varepsilon_1+\varepsilon_2}
		e^{-Q_\parallel\abs{z-z'}}
		&,\quad\text{sgn}(z) \neq \text{sgn}(z')\label{eq:greenfunctionstatic}
		\\
		-\frac{1}{2 Q_\parallel}
		e^{-Q_\parallel\abs{z-z'}}
		-\frac{1}{2Q_\parallel}
		\frac{\varepsilon_2-\varepsilon_1}{\varepsilon_1+\varepsilon_2}
		e^{-Q_\parallel\abs{z+z'}}
		&,\quad z,z' < 0. 
	\end{cases}
\end{align}
\end{widetext}
This Green's function is derived for the quasi-static case of $c\rightarrow\infty$ which is caused by the proximity of the AuNP and TMDC. Hence, to a good approximation, we can neglect radiative interactions. We find that the Green's function is defined piecewise depending on the position of the source $z'$ and the observation location $z$.

\section{Eigensystem}
\label{sec:eigensystem_appendix}

Our numerical implementation enables a comprehensive investigation of various parameter configurations for our system, which we will explore in the following sections. First, we analyze the dispersion relation for varying background permittivity $\varepsilon_2$ in Sec.~\ref{sec:dispersion}. We then study the spectral eigenvalue distribution in Sec.~\ref{sec:eigenvalues_appendix}. In Sec.~\ref{sec:aspect_ratio}, we analyze the dependence of the lowest eigenvalue on the nanoparticle aspect ratio to quantify the interaction strength. Finally, we investigate the corresponding eigenvectors (plexcitonic wavefunctions) in Sec.~\ref{sec:eigenvectors_appendix}.

\subsection{Dispersion}
\label{sec:dispersion}

\begin{figure}[h!]
    \includegraphics[width=\linewidth]{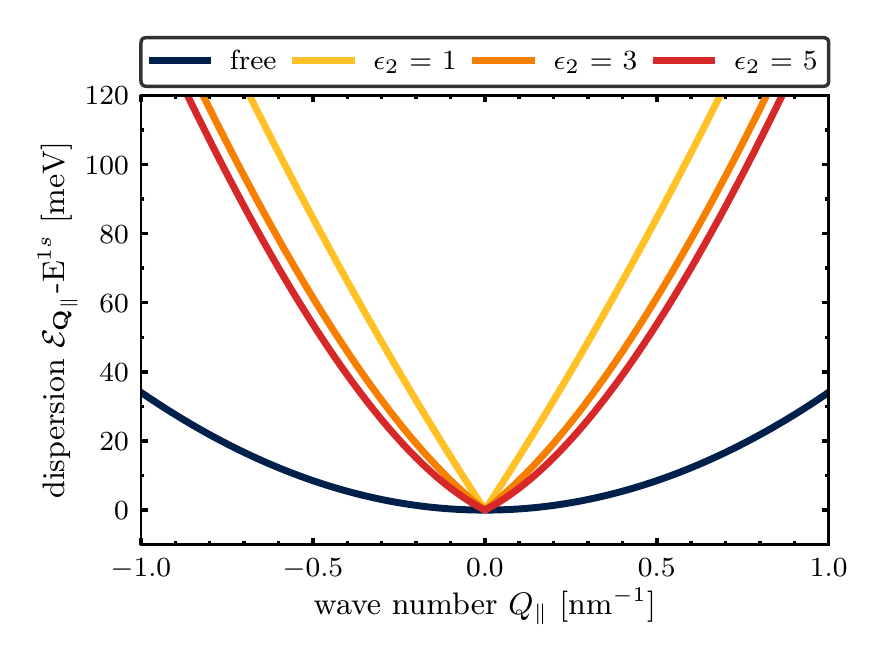}
    \caption{ \textbf{Dispersion relation.} The interacting dispersion  $\mathcal{E}_{\vb{Q}_\p}^V$, cf.~Eq.~\eqref{eq:dispersion_plexcitonic}, is plotted for varying background permittivity $\varepsilon_2$ from yellow to red. One can see that the dispersion interpolates between linear and parabolic behavior. The free  dispersion  $\mathcal{E}_{\vb{Q}_\p}^U$ [cf.~Eq.~\eqref{eq:dispersion_excitonic}] is plotted for reference in (dark) blue.}
    \label{fig:dispersion_comparison}
\end{figure}

The eigenvalues of the plexcitonic eigenvalue equation, Eq.~\eqref{eq:eigenvalue}, quasi-continuously distribute among the dispersion relation $\mathcal{E}_{\vb{Q}_\p}^V$. Only the negative eigenvalues differ from this distribution. Thus, we begin studying this dispersion relation for varying background permittivity $\varepsilon_2$ in Fig. \ref{fig:dispersion_comparison}.

The purely excitonic dispersion $\mathcal{E}_{\vb{Q}_\p}^U$ is not affected by changing the background permittivity $\varepsilon_2$ while the plexcitonic dispersion  $\mathcal{E}_{\vb{Q}_\p}^V$ is conical \cite{qiu2015nonanalyticity} for small $\varepsilon_2$ but becomes predominantly parabolic for larger $\varepsilon_2$.
%
\subsection{Eigenvalues}
\label{sec:eigenvalues_appendix}

\begin{figure}[h!]
    \includegraphics[width=\linewidth]{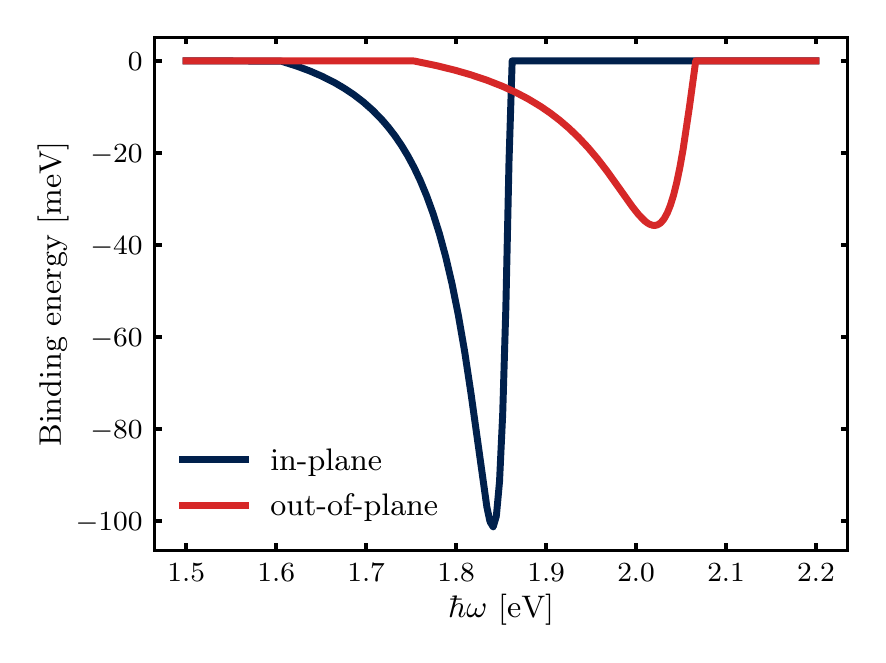}
    \caption{\textbf{Lowest eigenvalues} The lowest eigenvalue of the plexcitonic eigenvalue equation, cf.~Eq.~\eqref{eq:eigenvalue}, highly depends on the choice of the spectral position of the excitonic 1s resonance. Here, the lowest eigenvalue for interacting via the in-plane/out-of-plane axis is given. 
    }
\label{fig:eigenvalues_spectrum}
\end{figure}

In the following analysis, we examine the plexcitonic eigenvalues that deviate from the dispersion and become negative. This occurs when the TMDC exciton and AuNP plasmon are in resonance. Here, we focus on the distribution of the lowest eigenvalues, which we present in Fig.~\ref{fig:eigenvalues_spectrum}. Specifically, we explore two distinct distributions based on the interaction through the in-plane polarizability axis $\alpha_\p$ of the nanoparticle and the out-of-plane axis $\alpha_z$.

As anticipated, one can clearly distinguish the distributions of eigenvalues resulting from in-plane and out-of-plane interaction, respectively. We observe spectral ranges where negative eigenvalues are absent, either due to interaction via a single axis or in a narrow spectral range for both axes. In the absence of negative eigenvalues, the individual components are out of resonance, which prevents attractive interactions. However, when the components are in resonance, we detect negative eigenvalues, which we interpret as the binding energy of the exciton in the potential induced by the AuNP. Notably, at certain spectral positions, these binding energies amount to several tens of meV.
%
\subsection{AuNP aspect ratio}
\label{sec:aspect_ratio}

\begin{figure}[h!]
    \centering
    \includegraphics[width=\linewidth]{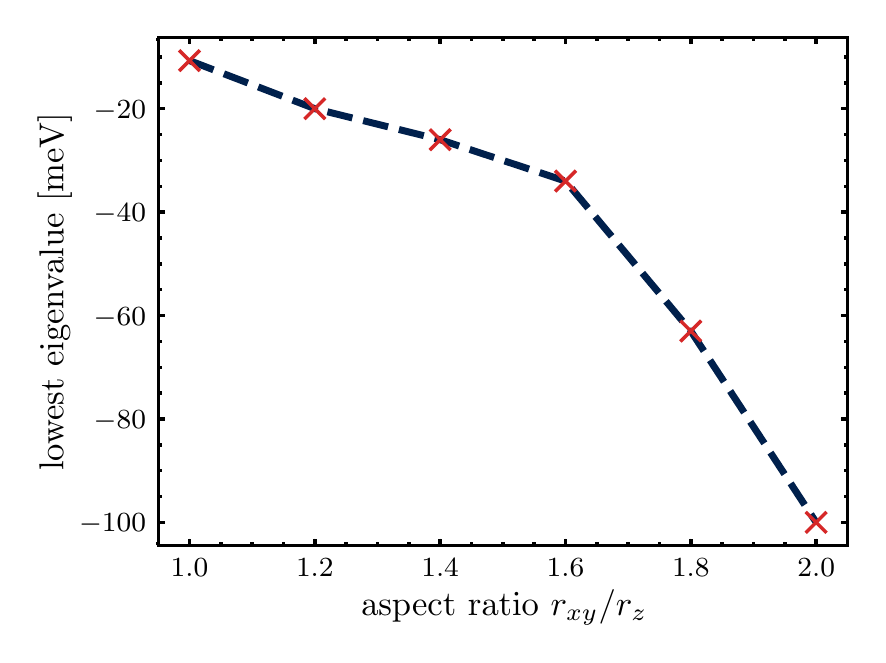}
    \caption{\textbf{Influence of Aspect Ratio on Lowest Eigenvalue.} By varying the aspect ratio, we can adjust the effective distance between the AuNP and the TMDC layer for a fixed volume and thus modify the interaction strength. The plot displays the dependence of the lowest eigenvalue on the aspect ratio while maintaining a fixed volume.}
    \label{fig:aspectratio}
\end{figure}

In our numerical analysis presented in the main article, we consider spheroidal nanoparticles with an aspect ratio $\flatfrac{r_{xy}}{r_z}=2$. This choice of aspect ratio allows us to increase the nanoparticle volume while simultaneously decreasing the effective separation between the nanoparticle and TMDC layer. To investigate the influence of the aspect ratio $\flatfrac{r_{xy}}{r_z}$ on the coupling strength and quantify its effects, we recalculate the lowest eigenvalue for varying aspect ratio while keeping the nanoparticle volume fixed and position it right at the interface. The real part of the lowest eigenvalue as a function of aspect ratio is plotted in Fig.~\ref{fig:aspectratio}.

Our analysis shows that the binding strength increases as the aspect ratio of the spheroidal nanoparticle increases, resulting in a more negative lowest eigenvalue. This trend arises from the exponential dependence of the interaction strength on the separation, while the increase in volume with respect to the semi-axis in the out-of-plane direction is only linear. Therefore, we demonstrate that oblate spheroids exhibit a stronger interaction compared to a sphere of equal volume.

\subsection{Eigenvectors}
\label{sec:eigenvectors_appendix}

In Section~\ref{sec:eigenvectors}, we presented the probability density resulting from interaction via the in-plane components of the AuNP. This probability density is ring-shaped around the origin due to dipole-dipole interaction. For completeness, we now provide the probability density corresponding to interaction via the out-of-plane component.

\begin{figure}[h!]
    \centering
    \includegraphics{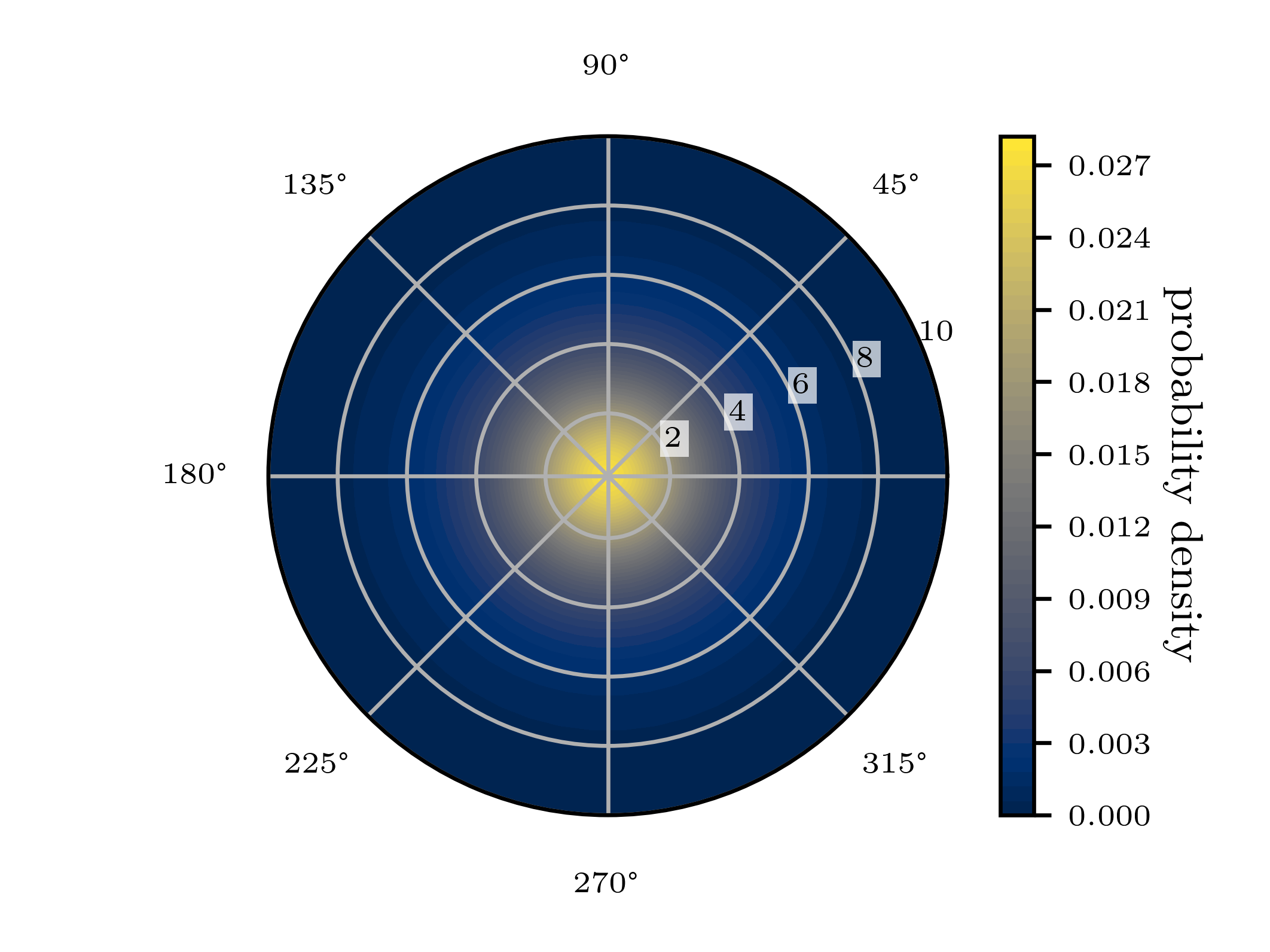}
    \begin{picture}(0,0)
    \put(-130,0){\includegraphics[width = .5\linewidth]{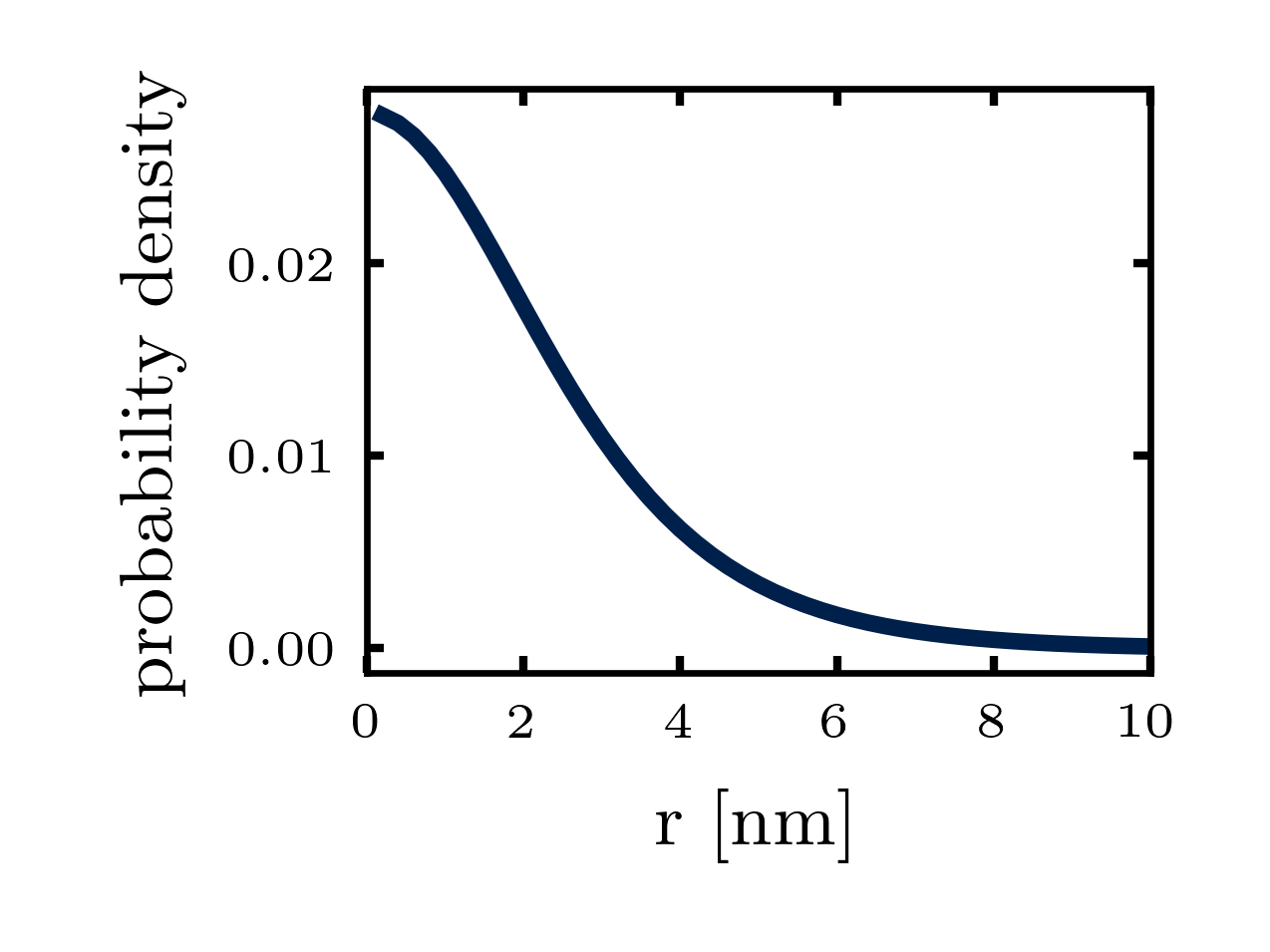}}
    \end{picture}
\caption
{
    \textbf{Probability density in real space.}  Here, we plot the probability corresponding to interaction via the out-of-plane axis of the AuNP. The inset shows the radial profile of the probability density.
}
\label{fig:wavefunction_appendix}
\end{figure}


In contrast to the probability density resulting from in-plane interaction, the probability density for out-of-plane interaction exhibits a Gaussian distribution centered around the origin of the AuNP position. This outcome can also be derived from the minimization of the dipole-dipole potential for dipoles that are perpendicular to one other.\\

\section{Macroscopic quantities}
\label{sec:macroscopic_quantities}

In this section, we will shift our focus from the microscopic quantities discussed in the previous section to observable macroscopic quantities that can be derived from our calculations. Specifically, in Sec.~\ref{sec:TMDC_polarization_appendix}, we will examine the circularly polarized components of the macroscopic TMDC polarization in real space. Additionally, in Sec.~\ref{sec:gold_polarization}, we will provide an analytical expression for the AuNP polarization, which serves as a source in the Green's function approach used to calculate the electric field. In Sec.\ref{sec:rabisplitting_appendix}, we will analyze the parameter dependence of the Rabi splitting that was obtained in Sec.~\ref{sec:strongcoupling}.

\begin{figure*}[htb]
\subfloat
{
    \includegraphics[width=.49\textwidth]{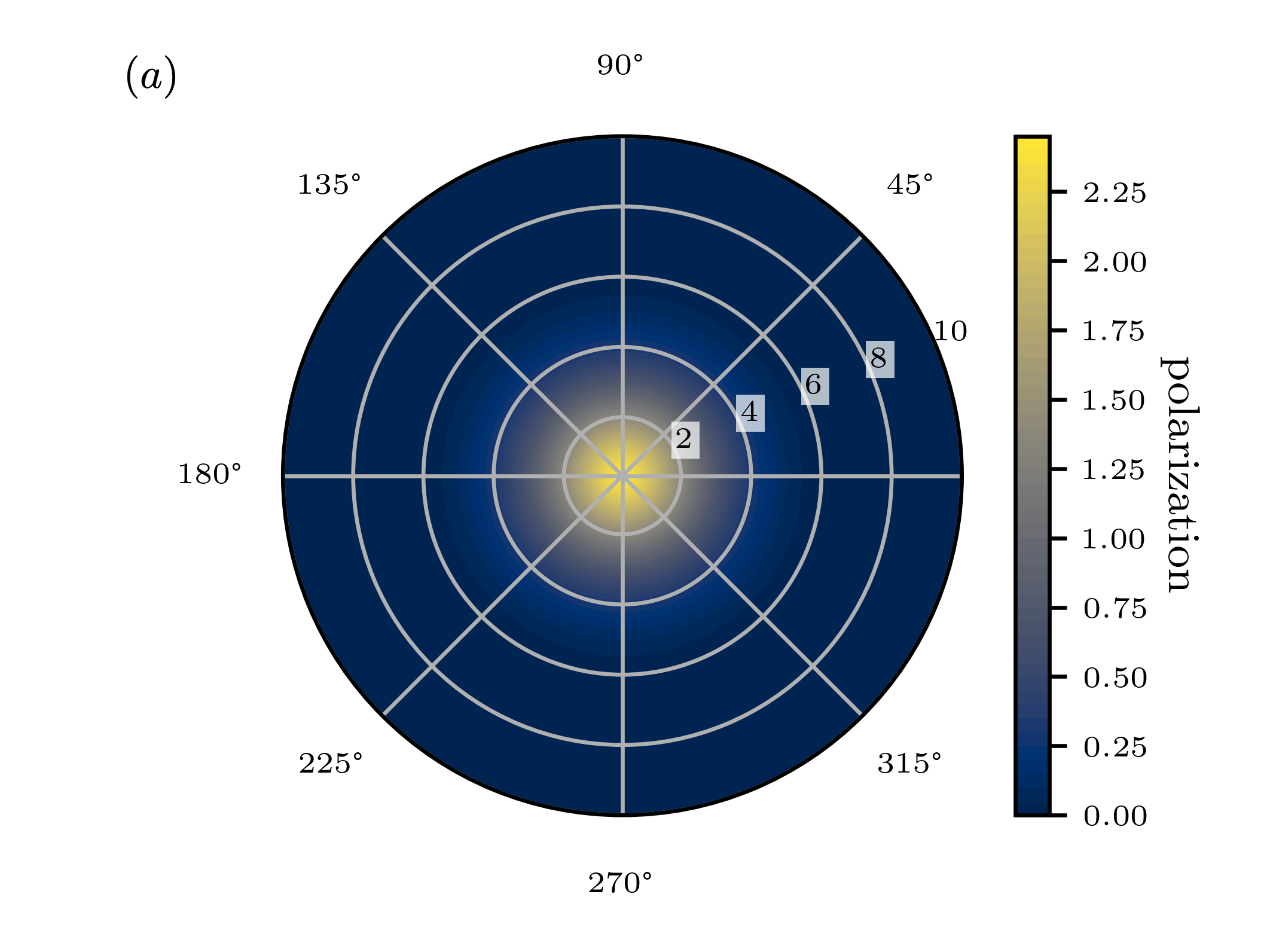}
}
\subfloat
{
    \includegraphics[width=.49\textwidth]{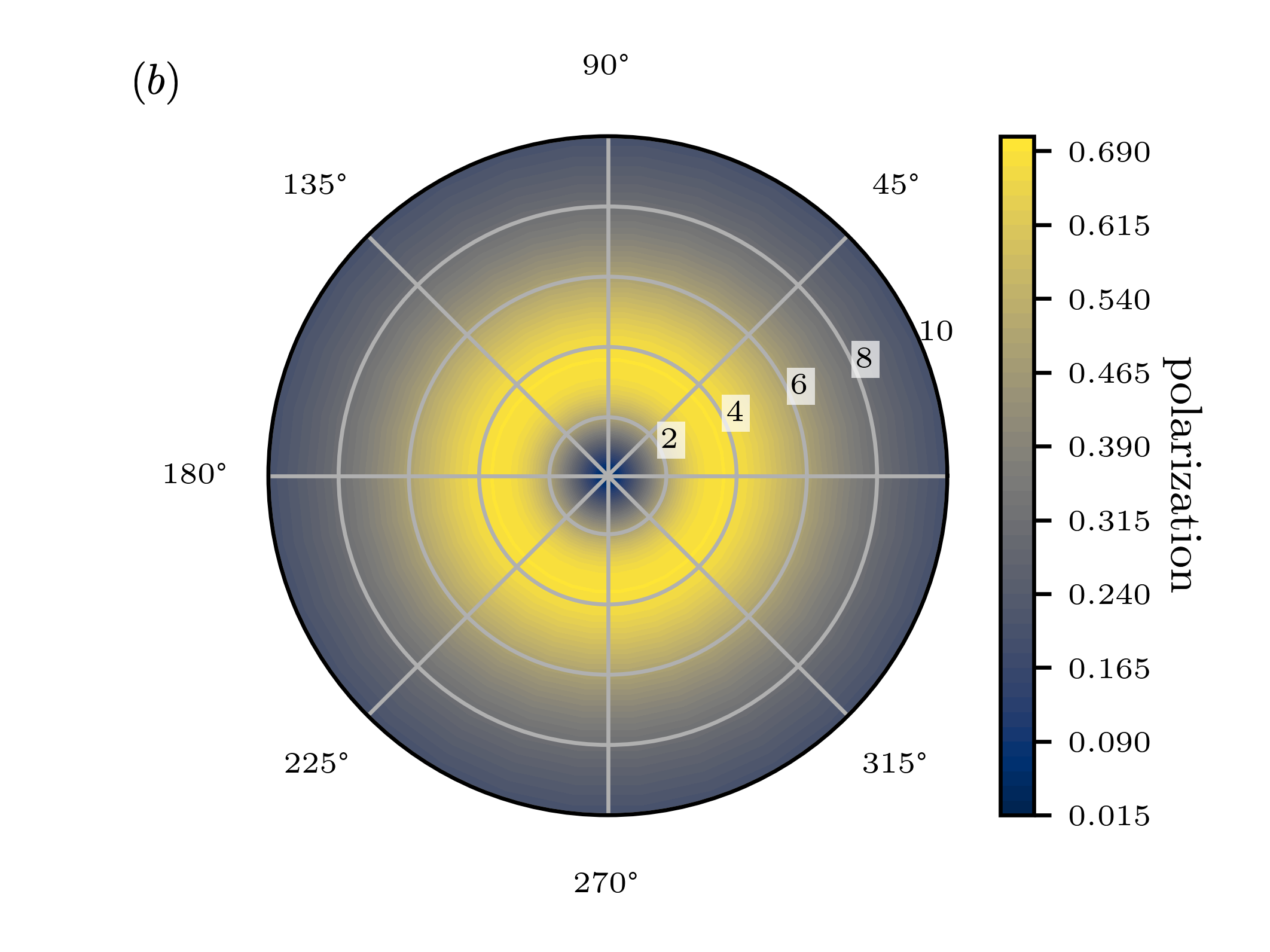}
}
\caption
{
    \textbf{Macroscopic TMDC polarization.} We plot $\sigma^+$ polarization in (a) and $\sigma^-$ polarization in (b) for the case of $\sigma^+$ excitation. In (a), the $\sigma^+$ contribution is centered around the origin, while in (b), the $\sigma^-$ contribution forms a ring around the origin. The full polarization, shown in Fig.~\ref{fig:polarization}, is obtained as the sum of both contributions.
}
\label{fig:polarization_appendix}
\end{figure*}

\subsection{TMDC polarization}
\label{sec:TMDC_polarization_appendix}

In Sec.~\ref{sec:localization}, we presented the full TMDC polarization following excitation with a $\sigma^+$ pulse. In this section, we analyze the individual $\sigma^+/\sigma^-$ components of the polarization after excitation with a $\sigma^+$ pulse. The results are analogous for $\sigma^-$ excitation.

%

Our analysis reveals two distinct shapes for the spatial distribution of the macroscopic TMDC polarization. The absolute value of the $\sigma^+$ polarization is Gaussian distributed and centered around the origin. For the absolute value of the $\sigma^-$ polarization, a ring-shaped feature is observed, similar to the probability density investigated in Sec.\ref{sec:eigenvectors}, with vanishing polarization at the origin.
These findings suggest that the selection rules are modified in the electric near-field \cite{knorr1999theory}, enabling the excitation of oppositely polarized light. Furthermore, we interpret our results as indicating that polarization of the same direction is primarily induced by the external field that scatters off the AuNP, whereas polarization of the opposite direction mostly originates from the dipole-dipole interaction between the TMDC exciton and AuNP plasmon, reproducing the shape of the probability density from Sec.~\ref{sec:eigenvectors}.

\subsection{Gold polarization}
\label{sec:gold_polarization}

In Sec.\ref{sec:strongcoupling}, we use Eq.~\eqref{eq:field} to propagate the material polarizations to the surroundings. This material polarization includes an effective polarization originating from the TMDC, whose Fourier transformed version we provided in Eq.~\eqref{eq:TMDCpolarization}. Here, we will now provide the AuNP polarization in the mixed basis $(\vb{Q}_\p, z ;\,\omega)$, that is only defined at the spatial position of the AuNP,
\begin{widetext}
\begin{align}
\label{eq:goldpolarization_full}
    \vb{P}_{\vb{Q_\p}}^{\text{AuNP},\pm}(z_{\text{pl}};\omega)  = 
    \frac{1}{(2\pi)^2}&\int \dd^2 \vb{Q_\p'}
    \mqty( \alpha_\p E_{\vb{Q_\p'}}^+(z_{\text{pl}})\\ \alpha_\p E_{\vb{Q_\p'}}^-(z_{\text{pl}})\\ \alpha_{z} E_{\vb{Q_\p'}}^z(z_{\text{pl}}))\\\nonumber
    +\frac{\abs{d}^2\abs{\varphi_0}^2}{2}&\sum_\lambda \frac{1}{E+E^\lambda -\hbar\omega -i\gamma}
    \frac{1}{(2\pi)^2}\int \dd^2\vb{Q_\p'}\psi_{\vb{Q_\p'}}^R \frac{{Q_\parallel'}^2}{\varepsilon_0\varepsilon_1}G_{\vb{Q_\p'}}^{\text{st}}(z_{\text{pl}},z_{\text{ex}})\mqty(e^{-i\phi'}\alpha_\p\\e^{i\phi'}\alpha_\p\\ i \alpha_{z}) 
    \\\times
        \frac{\alpha_\p}{(2\pi)^2}\int\dd^2\vb{Q_\p''}\qty(\psi_{\vb{Q_\p''}}^L)^*
        &\biggl[\qty(e^{i\phi''}E_{\vb{Q_\p''}}^+(z_{\text{ex}})+e^{-i\phi''}E_{\vb{Q_\p''}}^-(z_{\text{ex}}))
        +\frac{Q_\parallel^2}{\varepsilon_0\varepsilon_2} G_{\vb{Q_\p''}}^{\text{st}}(z_{\text{ex}},z_{\text{pl}})
            \qty(e^{i\phi''}E^+(\vb{r}_{\text{pl}})
                +e^{-i\phi''}E^-(\vb{r}_{\text{pl}}))
    \biggr]\nonumber .
\end{align}
\end{widetext}
The polarization of the system is composed of three contributions. The first one is the dipole response of the AuNP, which is determined by its polarizability $\boldsymbol{\alpha}$ and the external electric field at the AuNP position $\vb{E}_{\vb{Q_\p'}}(z_{\text{pl}})$. The second contribution arises from the interaction of the external electric field with the TMDC layer, which is then mediated to the AuNP. The third contribution arises from the effective self-interaction of the plasmon, mediated via the TMDC layer.

\subsection{Rabi splitting}
\label{sec:rabisplitting_appendix}

\begin{figure*}[htb]
    \centering
    \includegraphics[width = \linewidth]{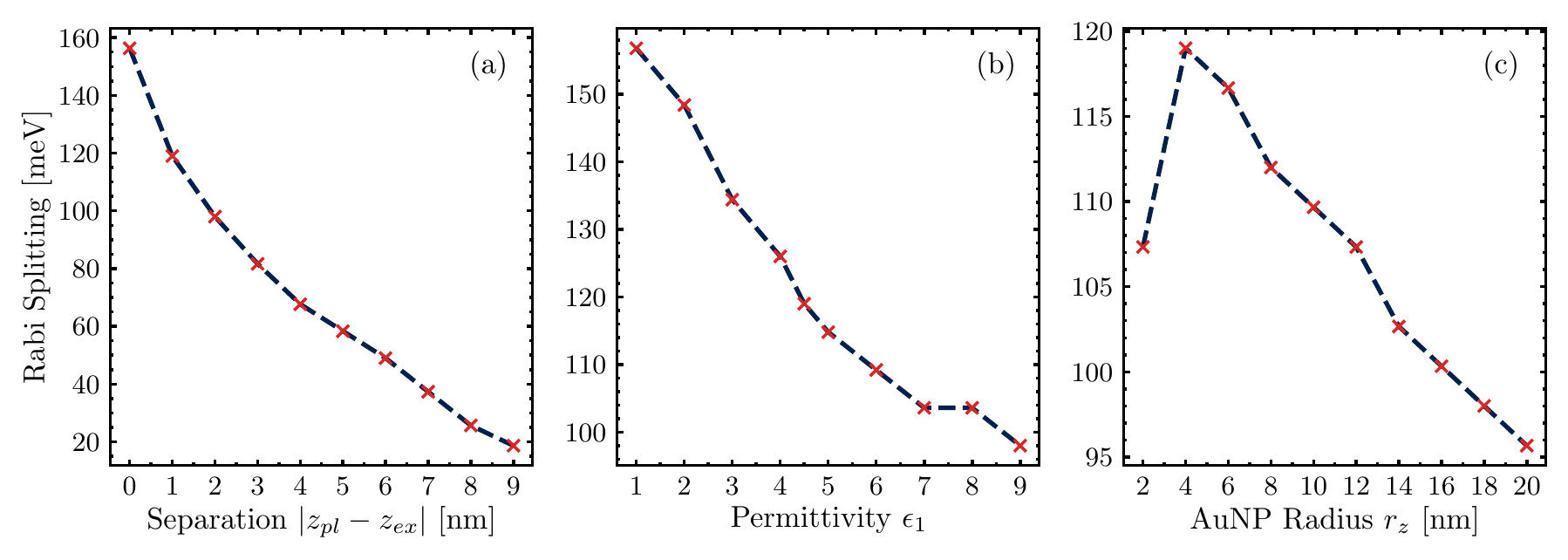}
    \caption{\textbf{Rabi Splitting:} The graphs display the dependence of the Rabi splitting over a parameter range that results in strong coupling. Fig.~(a) illustrates that an increase of the separation of TMDC and AuNP results in a decrease of the splitting. Fig.~(b): For an increasing background permittivity $\varepsilon_1$ in the upper half space, the Rabi splitting decreases. Fig.~(c): For an increasing AuNP radius $r_z$, the Rabi splitting reaches a maximum at $\approx\unit[4]{nm}$. This behavior  qualitatively agrees with the dependence of the lowest eigenvalue on the AuNP radius.}
    \label{fig:parameter}
\end{figure*}

In Fig.~\ref{fig:strongcoupling}, we observed that our system reveals a Rabi splitting of several tens of meV and thus clearly operates in the strong coupling regime. The Rabi splitting can be tuned via various system parameters, which we analyze individually to understand their impact on $\Omega$ in Fig.~\ref{fig:parameter}.

In our study, we observe a decrease in the Rabi splitting as the TMDC/AuNP spacing decreases, consistent with the findings in Ref.~\cite{carlson2021strong}. This decrease can be attributed to the significant reduction in the interaction strength as the separation between the materials increases.

Furthermore, we investigate the impact of increasing background permittivity $\epsilon_1$ in the upper half-space on the Rabi splitting. We find that as the background permittivity increases, the Rabi splitting decreases. This can be interpreted as the enhanced screening effect resulting from the increased background permittivity, which weakens the overall interaction. 

Lastly, we consider the influence of particle radius on the Rabi splitting, which exhibits a scaling behavior similar to the dependence of the plexcitonic eigenvalues $E^\lambda$ on the radius. Increasing the radius leads to a cubic increase in volume, enhancing the interaction strength. However, this is counteracted by the increase in effective separation, leading to an exponential decrease in the interaction strength.

\bibliography{references}
\end{document}